\documentclass[11pt]{article}
\newcommand{\Comment}[1]{{}}
\usepackage{amsfonts,amsthm,amsmath,amssymb,slashed}
\usepackage[textwidth = 430 pt, textheight = 630 pt]{geometry}
\usepackage{color}

\definecolor{MyDarkBlue}{rgb}{0.15,0.15,0.45}
\usepackage[linktocpage=true]{hyperref}
\hypersetup{
colorlinks=true,
citecolor=MyDarkBlue,
linkcolor=MyDarkBlue,
}

\usepackage{graphicx}
\usepackage{cite}

\newcommand\ignore[1]{}
\def\one{{\,\hbox{1\kern-.8mm l}}}

\def\a{\alpha}\def\b{\beta}

\def\d{\partial}

\newcommand{\Cset}{{\,\,{{{^{_{\pmb{\mid}}}}\kern-.45em{\mathrm C}}}}}

\newcommand{\be}{\begin{equation}}
\newcommand{\bea}{\begin{eqnarray}}

\newcommand{\ee}{\end{equation}}
\newcommand{\eea}{\end{eqnarray}}

\newcommand{\non}{\nonumber \\}
\newcommand{\CR}{\non\cr}
\newcommand{\pa}{\partial}

\def\pa{\partial}

\begin{document}

\renewcommand{\thefootnote}{\fnsymbol{footnote}}

\makeatletter
\@addtoreset{equation}{section}
\makeatother
\renewcommand{\theequation}{\thesection.\arabic{equation}}

\rightline{}
\rightline{}




\begin{center}
{\LARGE \bf{\sc CSBIon - a charged soliton of the 3-dimensional CS + BI Abelian gauge theory}}
\end{center} 
 \vspace{1truecm}
\thispagestyle{empty} \centerline{
{\large \bf {\sc Horatiu Nastase${}^{a}$}}\footnote{E-mail address: \Comment{\href{mailto:horatiu.nastase@unesp.br}}
{\tt horatiu.nastase@unesp.br}}
{\bf{\sc and}}
{\large \bf {\sc Jacob Sonnenschein${}^{b,c}$}}\footnote{E-mail address: \Comment{\href{mailto:cobi@tauex.tau.ac.il}}{\tt cobi@tauex.tau.ac.il}}
                                                        }

\vspace{.5cm}


\centerline{{\it ${}^a$Instituto de F\'{i}sica Te\'{o}rica, UNESP-Universidade Estadual Paulista}} 
\centerline{{\it R. Dr. Bento T. Ferraz 271, Bl. II, Sao Paulo 01140-070, SP, Brazil}}
\vspace{.3cm}
\centerline{{\it ${}^b$School of Physics and Astronomy,}}
\centerline{{\it The Raymond and Beverly Sackler Faculty of Exact Sciences, }} \centerline{{\it Tel Aviv University, Ramat Aviv 69978, Israel}}
\vspace{.3cm}
\centerline{{\it ${}^c$ Simons Center for Geometry and Physics,}} 
\centerline{{\it SUNY, Stony Brook, NY 11794, USA }}

\vspace{1truecm}

\thispagestyle{empty}

\centerline{\sc Abstract}

\vspace{.4truecm}

\begin{center}
\begin{minipage}[c]{380pt}
{\noindent
In this paper, we construct 
 a charged soliton with a finite energy and no delta function 
source in a pure Abelian gauge theory. Specifically, we first consider the 3-dimensional Abelian gauge 
theory, with a
Maxwell term  and a level $N$ CS term.  We find a static solution that carries charge $N$, 
angular momentum $\frac{N}{2}$ and whose radius is $N$ independent. However, this solution has a divergent energy.
In analogy to  the  replacement of  the 4 dimensional 
 Maxwell action with the BI action,  which renders the classical energy of a point charge finite, for the 3 dimensional 
 theory which includes a CS term such a replacement leads to a finite energy for the solution of above. 
 We refer to this soliton as a  CSBIon
solution, representing a finite energy version of the fundamental (sourced) charged electron of Maxwell 
theory in 4 dimensions. In 3 dimensions the BI+CS action has a static charged solution with finite energy and 
no source, hence a soliton solution. 
The CSBIon, similar to its Maxwellian predecessor, has a  charge $N$, angular momentum proportional 
to $N$ and  an $N$-independent radius.
 We also present other nonlinear modifications of Maxwell
theory that admit  similar solitons. 
The CSBIon may be relevant in various  holographic scenarios. In particular, it may describe a D6-brane 
wrapping an $S^4$ in a compactified D4-brane background. 
We believe that the CSBIon may play a role  in  condensed matter systems  in 2+1 dimensions like graphene sheets.

}
\end{minipage}
\end{center}

\vspace{.5cm}

\setcounter{page}{0}
\setcounter{tocdepth}{2}

\newpage

\tableofcontents
\renewcommand{\thefootnote}{\arabic{footnote}}
\setcounter{footnote}{0}

\linespread{1.1}
\parskip 4pt



\section{Introduction}

Classical solutions of quantum field theories with finite energy are physically very important and are rare. 
In gauge theories there are  certain finite energy solutions with some finite charge, usually topological in nature, though not only 
(for instance, consider the Q-ball solution \cite{Coleman:1985ki}). 
In the case of nonabelian gauge theories, one can have topological soliton 
solutions involving the gauge fields only, for instance the BPST instanton solution\cite{Belavin:1975fg}, 
though in that case the solution only 
exists in Euclidean signature. If one adds matter, specifically scalars, there are more soliton solutions possible, like 
the 't Hooft monopole in the 3+1 dimensional nonabelian case\cite{tHooft:1974kcl}, and the Nielsen-Olesen vortex in 
2+1 dimensional Abelian-Higgs theory\cite{Nielsen:1973cs}. One can also have finite energy solutions that  
are sourced by a delta function, like the BIon solution, invented by Born and Infeld\cite{Born:1934gh} 
in order to describe the electron 
as a finite energy solution with a delta function source. 

But until now, to our knowledge, there were no soliton solutions in pure abelian gauge theory. In this paper, we  
first  derive  a static solution of the Maxwell + level $N$  CS theory. This explicit solution has a charge $N$, 
 angular momentum $N/2$ and a radius which is $N$ independent. However, it has a divergent energy and a delta 
 function source. 
 We cure both problems by uplifting the system  into a BI + CS one.
We refer to the corresponding soliton solution as the CSBIon.  For that case were not able to derive 
 an analytic explicit solution, but  we show that indeed it has finite energy, and charge, angular 
 momentum and radius similar to those of the predecessor Maxwell + CS theory, but no delta function source.   Moreover, the
electric charge associated with the solution does not arise from a topological number. 

The Maxwell + CS electromagnetism  in 2+1 dimensions has  many  applications to condensed matter physics.
 These are described in the reviews \cite{ Wen:1995qn, Zee:1995avy, Witten:2015aoa}
 and in references therein. Probably in a similar manner one can consider applications of the BI + CS action to solid states systems.
In particular a phenomenological description of the dynamics of the graphene sheets in terms of a DBI action was proposed in \cite{Babington:2010vv}. The CSBIon may be a source outside of the sheet. 

Gauge field theories, abelian and non-abelian, described by  an action built of  BI and  CS terms,  
are very common  on the worldvolumes of D-branes. As such they show up in various string and holographic models.
An example of  such an abelian gauge  theory in three dimensions is associated with a D6-brane that resides in the 
background of  compactified D4-branes  and wraps an $S^4$. This model has been suggested \cite{BigCot} as the holographic 
dual of the proposal to describe an $N_f=1$ baryon in terms of a  quantum Hall droplet\cite{Komargodski:2018odf}.

The paper is organized as follows. The next section is devoted to the
motivation  for this work and to  a comparison with BIon solution in 4 dimensions. In section  3  we derive 
solutions of the Maxwell + CS action. First we derive the basic static solution and compute its  classical energy, 
angular momentum and radius. We then derive a solution with finite energy for the case where  the origin is 
encircled by a conducting circle and a time dependent solution. In section 4  we uplift the Maxwell term to a 
BI one. We write down the equations of motion and the constitutive relations. We analyze the structure of 
the solution and conclude that it has to have  finite energy and  charge and angular momentum that are 
linear with $N$ and radius which is independent of it. Next we describe certain  ModMax generalizations. 
In the next section we summarize, conclude and write down several open questions. The paper includes 
also three appendices. In the first we describe a non-relativistic BI-type model, followed in the second by a 
relativistic one. We then present 4 attempts of approximating the exact solution in the third one.

\section{Motivation and comparison with BIon solution in 4 dimensions}

As motivation for our work, we can take the point of view of the formal theoretical physicist, and simply 
look for an answer to a mathematical physics question: can we find in Abelian gauge theory  a finite energy soliton solution, 
which is not sourced by a delta function?

In 4 dimensions, the BIon solution to the BI action \cite{Born:1934gh} 
(modification of Maxwell electromagnetism)
has a finite energy, which is why Born and Infeld constructed it. But it is also sourced by a delta function, 
so as to be able to be identified with a finite field energy version of the electron. At $r\rightarrow \infty$, the 
BIon solution becomes the regular Maxwell electron, so $\vec{E}\propto 1/r^2$, which 
gives a finite energy at infinity, since ${\cal E}\sim 4\pi \int r^2 dr \vec{E}^2/2\sim \int dr/r^2$, while at 
$r\rightarrow 0$, the BIon modification keeps $\vec{E}$ finite. 

But the BIon is necessarily sourced, since $\vec{\nabla}\cdot \vec{D}\equiv 4\pi \tilde \rho_f$, with 
$\tilde \rho_f$ the {\em free}, 
or external, charge density, which is found to be $q\delta^3(r)$. There are no static solutions that are 
finite energy and not sourced, either in Maxwell or in BI theory.

 In Maxwell theory
(see \cite{Ranada:1989wc,ranada1990knotted,Sircar:2016old}) and in its BI generalization 
\cite{Alves:2017zjt,Alves:2017ggb}, there are  time-dependent knotted solutions with non-trivial topological charges. 

So it is natural to look to 3 dimensions, and see if we can find something there. But in 3 dimensions, 
even the regular Maxwell electron has $\vec{E}\propto 1/r$, so a diverging energy at infinity, since now
${\cal E}\sim 2\pi \int rdr \vec{E}^2/2\sim \int dr/r$. So one needs to consider a modification of Maxwell
theory {\em at large distances, or small energies (in the IR)}. Luckily, in 3 dimensions we have the 
CS term that we can add, and will dominate in the IR. 

We can now ask: can we find such an action, of Maxwell + CS, or BI + CS in a physical system? 
The answer for BI+ CS is in the affirmative, as follows. 

Consider the D4-brane holographic system, or the doubly-Wick rotated nonextremal
D4-brane (Witten model) with a large $N$ number of D4-branes, and consider 
a D6-brane wrapping the transverse $S^4$ in it, and the other 3 directions being parallel to the 
D4-brane. The CS term on the D6-brane will contain a nontrivial term of the type $\int A\wedge dA
\wedge F_{(4)}$, and since on the transverse sphere $F_{(4)}\sim N\epsilon_{(4)}$, we obtain 
on the 3 directions common to the D4- and D6-brane an Abelian gauge theory term 
\be
S_{\rm CS+BI}=S_{\rm BI}+\frac{N}{2\pi}\int d^3x \epsilon^{\mu\nu\rho}A_\mu\d_\nu A_\rho.
\ee

But, before we continue, we will review the 4-dimensional BIon solution.

The 4-dimensional BI action is 
\be
{\cal L}(b;\vec{E},\vec{B})=b^2\left[1-\sqrt{1+F-G^2}\right]\;,\label{LagDBI}
\ee
where $b$ is the dimensional parameter, of dimension 2, that defines the theory, and
\be
F=\frac{1}{b^2}(\vec{B}^2-\vec{E}^2)=\frac{1}{2b^2}F_{\mu\nu}F^{\mu\nu}\;,\;\;
G=\frac{1}{b^2}\vec{E}\cdot\vec{B}=-\frac{1}{4b^2}F_{\mu\nu}\tilde F^{\mu\nu}\;,
\ee
with $\tilde F^{\mu\nu}=\frac{1}{2}\epsilon^{\mu\nu\rho\sigma}F_{\rho\sigma}$.

As always in nonlinear electromagnetism theories, be it inside a material, or in vacuum, 
we define the objects
\be
\vec{H}=-\frac{\d {\cal L}}{\d \vec{B}}=\frac{\vec{B}-G\vec{E}}{\sqrt{1+F-G^2}}\;,\;\;
\vec{D}=\frac{\d {\cal L}}{\d \vec{E}}=\frac{\vec{E}+G\vec{B}}{\sqrt{1+F-G^2}}\;,
\ee
the above $H(E,B)$ and $D(E,B)$ being constitutive relations for the material, or the vacuum theory.

In terms of $\vec{E},\vec{D},\vec{B},\vec{H}$, the Maxwell equations without sources have form
\bea
\vec{\nabla}\times \vec{E}=-\frac{1}{c}\d_t \vec{B}\;, && \vec{\nabla}\cdot \vec{B}=0\;,\cr
\vec{\nabla}\times \vec{H}=\frac{1}{c}\d_t \vec{D}\;, && \vec{\nabla}\cdot \vec{D}=0.\label{Mxmedium}
\eea

In the presence of sources, one has
\be
\vec{\nabla}\cdot\vec{D}=\tilde \rho_{\rm ext}\;,
\ee
which contains only the {\em external} (or free) charge density $\tilde \rho_{\rm ext}$ (or $\tilde\rho_f$), which means
delta function sources, introduced as an extra term in the Lagrangian of the type $\int \tilde \rho_{\rm ext}
A_0$, whereas we also have
\be
\vec{\nabla}\cdot \vec{E}=\frac{\tilde\rho}{\epsilon_0}\;,
\ee 
but here in $\tilde \rho$ we also have charges due to the polarization 
of the material, or in this case, of the vacuum, leading as usual to the fact that
 this {\em total} charge density is spread out.
 
In 4 dimensions, the Hamiltonian is the Legendre transform of the Lagrangian over $\vec{E}
=F^{0i}=-\dot{\vec{A}}$ in the $A_0=0$ gauge, 
\be
{\cal H}=\vec{E}\vec{D}-{\cal L}=b^2\left[\frac{1+\frac{\vec{B}^2}{b^2}}{\sqrt{1+\frac{\vec{B}^2-\vec{E}^2}{b^2}-\left(\frac{\vec{B}\cdot
\vec{E}}{b^2}\right)^2}}-1\right]\;,\label{HamLeg}
\ee 
and since we can calculate that 
\bea
2s&\equiv & \vec{D}^2+\vec{B}^2=\frac{\vec{E}^2+\vec{B}^2\left(1+\frac{\vec{B}^2-\vec{E}^2}{b^2}\right)+2\frac{(\vec{E}\cdot\vec{B})^2}{b^2}}
{1+\frac{\vec{B}^2-\vec{E}^2}{b^2}-\left(\frac{\vec{B}\cdot\vec{E}}{b^2}\right)^2}\cr
p^2&\equiv &\vec{D}^2\vec{B}^2-(\vec{B}\cdot\vec{D})^2=
\frac{\vec{E}^2\vec{B}^2-\frac{(\vec{E}\cdot\vec{B})^2}{b^2}}{1+\frac{\vec{B}^2-\vec{E}^2}{b^2}
-\left(\frac{\vec{B}\cdot\vec{E}}{b^2}\right)^2}\;,
\eea
we can re-express it in terms of its natural variables, $\vec{D}$ and $\vec{B}$, as 
\be
{\cal H}(b;\vec{D},\vec{B})=b^2\left[\sqrt{1+\frac{2s}{b^2}+\frac{p^2}{b^4}}-1\right]=b^2\left[\sqrt{1+\frac{\vec{D}^2+\vec{B}^2}{b^2}
+\frac{\vec{D}^2\vec{B}^2-(\vec{D}\cdot\vec{B})^2}{b^4}}-1\right].\label{HamCorrect}
\ee 

The BIon is a purely electric solution ($\vec{B}=0$), sourced by a point charge, so 
$\tilde \rho_{\rm ext}=\Omega_{d-1}q \delta^d(\vec{r})$, where for later simplicity we took out a 
factor of $\Omega_{d-1}$, the volume of the unit sphere; for $d=3$, $\Omega_2=4\pi$. 

For the purely electric theory, the relevant constitutive relation becomes 
\be
\vec{D}=\frac{\vec{E}}{\sqrt{1-\vec{E}^2/b^2}}\;,
\ee
inverted as
\be
\vec{E}=\frac{\vec{D}}{\sqrt{1+\vec{D}^2/b^2}}=-\vec{\nabla}\phi.
\ee

Then, in 4 dimensions the equation of motion for the BIon solution becomes
\be
\frac{d}{dr}(r^2D_r)=4\pi q\delta^3(\vec{r})\;,
\ee
with solution 
\be
D_r=\frac{q}{r^2}\;,
\ee
so that 
\be
E_r=-\phi'(r)=\frac{q/r^2}{\sqrt{\frac{q^2}{b^2r^4}+1}}=\frac{qb}{\sqrt{b^2r^4+q^2}}.
\ee

As we see, at $r\rightarrow\infty$, the solution reduces to the Maxwell electron solution, and at 
$r\rightarrow 0$, $E/b\rightarrow 1$, the maximum allowed value, because of the 
square root $\sqrt{1-\vec{E}^2/b^2}$.

While $\vec{\nabla}\cdot \vec{D}=\tilde \rho_{\rm ext}=q\delta^3(\vec{r})$ is sourced by a point charge, 
the total charge is spread out, 
\be
\frac{\tilde \rho}{\epsilon_0}\equiv \vec{\nabla}\cdot \vec{E}=\frac{d}{dr}(r^2E_r)=\frac{d}{dr}
\frac{q}{\sqrt{\frac{q^2}{b^2r^2}+1}}=\frac{2q^3}{b^2r^5\left(\frac{q^2}{b^2r^4}
+1\right)^{3/2}}\;,
\ee
due to the ``polarization of the vacuum''.

The total field energy of the purely electric solution, the spatial integral of its Hamiltonian, 
\be
{\cal E}=\int d^3r b^2\left[\sqrt{1+\frac{\vec{D}^2}{b^2}}-1\right]=4\pi b^2\int_0^\infty 
r^2dr\left[\sqrt{1+\frac{q^2}{b^2r^4}}-1\right]\;,
\ee
is finite.

\subsection{3 dimensional BIon solution to BI theory}

We can repeat the same analysis for the 3-dimensional case. We now denote by the $\rho$ the 2-dimensional radial coordinate
(polar coordinate in the plane).

In 2+1 dimensions, the equation of motion for the BIon solution is (taking out a factor of $\Omega_1=2\pi$
as before),
\be
\frac{d}{d\rho}(\rho D_\rho)=2\pi q \delta^2(\vec{r})\;,
\ee
with solution 
\be
D_\rho=\frac{q}{\rho}\;,
\ee
so 
\be
E_\rho=\phi'=\frac{q/\rho}{\sqrt{\frac{q^2}{\rho^2b^2}+1}}=\frac{qb}{\sqrt{b^2\rho^2+q^2}}.
\ee

This integrates to 
\be
\phi=-q\int_0^\rho\frac{dx}{\sqrt{x^2+(q/b)^2}}=q\sinh^{-1}\frac{b\rho}{q}.
\ee

However, now the total field energy of the purely electric solution is 
\be
{\cal E}=\int d^2r b^2\left[\sqrt{1+\frac{\vec{D}^2}{b^2}}-1\right]=2\pi b^2\int_0^\infty 
\rho d\rho\left[\sqrt{1+\frac{q^2}{b^2\rho^2}}-1\right]\;,
\ee
and is log-divergent at $\rho\rightarrow \infty $ as $\int d\rho/\rho$, the same divergence as in the Maxwell 
case. Of course, at $\rho\rightarrow 0$ the energy is still finite.

\section{Solutions for Maxwell plus Chern-Simons in 3 dimensions}

In 3 dimensions, we can add a CS term, that will dominate over the Maxwell one (or a BI, reducing to 
Maxwell) at large distances, so in the IR. We analyze therefore the solutions of this system. 
\subsection {The basic static solution}
Consider then the Abelian Maxwell + CS term action at level $N$, that reads
\be
S_{\rm CS+Mx}= \int d^3 x \left[-\frac{1}{4 g^2}F_{\mu\nu}F^{\mu\nu} 
+ \frac{N}{2\pi} \epsilon^{\mu\nu\rho} A_\mu\pa_\nu A_\rho\right]\;,
\ee
where, since we have the CS term added to the Maxwell term, we have introduced also the coupling 
$g^2$ in front of the action. Then, as usual, $A_\mu$ has mass dimension 1, so $g^2$ has 
mass dimension 1.

The corresponding equation of motion is
\be
\pa_\nu F^{\nu\mu} + \lambda \epsilon^{\mu\nu\rho} F_{\nu\rho}=0\;,\label{coveom}
\ee 
where $\lambda= \frac{g^2 N}{2\pi}$ has dimension 1.

Explicitly, we have ($i=1,2$)
\bea\label{eqom}
\pa_i F^{i0}  + \lambda F_{12} & = & 0 \cr
\pa_0 F^{01} + \pa_2 F^{21} + \lambda F_{20} & = &  0\cr
\pa_0 F^{02} + \pa_1 F^{12} + \lambda F_{01} & = & 0.
\eea

We define, as usual, the magnetic field (in 3 dimensions, it is a scalar) $B\equiv F_{12}$, 
the electric field $E_i\equiv F_{i0}$. Consider a static solution ($\d_t \vec{E}=\d_t B=0$)
depending only on the radial coordinate $\rho$, the radial component of $\vec{E}$ 
denoted by $E$  and with $E'=\pa_\rho E$, $B'=\d_\rho B$.
Then  the equations of motion  take the form
\bea
\frac {E}{\rho} + E' & = &  \lambda B \cr
\pa_i B & = &  \lambda E_i \Rightarrow B'=\lambda E.
\eea

Combining the two, we obtain a single equation for $E$,
\be\label{BesselK}
 \rho^2 E''  + \rho E' - E( 1+ \lambda^2 \rho^2)= 0.
\ee 

Denoting $z\equiv\lambda\rho$, we obtain a modified Bessel equation in the variable $z$,
\be
z^2 \pa_z^2 E + z\pa_z E - E(1+ z^2) =0.
\ee

Thus the general solution for $E$ is 
\be 
E= \tilde a I_1[ \lambda\rho] + \tilde b K_1[ \lambda\rho]\;,
\ee
where $I_n[ \lambda\rho]$ and $K_n[ \lambda\rho]$  are the modified Bessel functions 
of first and second kind, and $\tilde a, \tilde b$ are arbitrary constants. 

Requiring on physical grounds that the field goes to zero at large $\rho$, so excluding the $I_1$ 
solution, we end up with the solution
\be
E = \tilde b K_1[ \lambda\rho ] \qquad B =  -\tilde b K_0[ \lambda\rho]. \label{solCS}
\ee

Near $\rho=0$, this solution becomes 
\be
E(\rho)\simeq \frac{\tilde b}{\lambda \rho}\;,\;\;\; 
B(\rho)\simeq \tilde b \ln\left(\frac{\lambda \rho}{2}\right).
\ee

We check that one of the equations of motion becomes near $\rho=0$
\be
B'\simeq \frac{\tilde b}{\rho}\simeq \lambda E\;,
\ee
so is satisfied near $\rho=0$, and the other becomes 
\be
E'+\frac{E}{\rho}\simeq \frac{\tilde b}{\lambda}\left(\frac{1}{\rho^2}
-\frac{1}{\rho^2}\right)\simeq \lambda 
B\;,
\ee
so is also satisfied, {\em but in leading order}, $1/\rho^2$ (if we keep higher orders in the expansions
of $E$ and $B$ in $\rho$, it is, of course, satisfied to all orders). 

In retrospect, to satisfy the two differential equations in leading order, we can propose the {\em ansatz}
that $E(\rho)\simeq \tilde b/(\lambda \rho)$, then find $B$ from $B'=\lambda E$, and then 
{\em check} that the remaining equation, $E'+E/\rho=\lambda B$, is satisfied {\em in leading order}.

Note, however, that the solution we found has a delta function source\footnote{We would like to thank Z. Komargodski for pointing this fact to us}. Similarly to what one does in 3+1 dimensions for the 
electron solution to pure Maxwell theory, 
we rewrite the 0 component of (\ref{coveom}) as
\be
\vec{\nabla}\cdot \vec{E}=\lambda B+C\delta^2(r)\;\label{deltasource}
\ee
with a free coefficient $C$, and integrate over an infinitesimal disk $D$ of radius $\epsilon$ in order to fix $C$. 
Using the Stokes theorem 
(Green-Riemann in 2 dimensions) to rewrite the left-hand side as $\int_C \vec{E}\cdot d\vec{l}$, we obtain 
\be
2\pi \frac{\tilde b}{\lambda}={\cal O}(\epsilon^2)+C\Rightarrow C=\frac{2\pi \tilde b}{\lambda}.
\ee

Note also that in this case, since we obtain a {\em linear} second order differential equation, with 
two independent solutions, we can also propose the other ansatz (corresponding to $E=I_1(\lambda 
\rho)$, which is excluded on physical grounds, as it blows up at infinity). 
Using the above rule, we would write (we introduce $D$ and $H$ for later use in the case of nonlinear electromagnetism 
theories, though here they are trivial, $D=E$, $B=H$)
\be
D=E\simeq A\rho +C\rho^3\Rightarrow B=H=\frac{1}{\lambda}\left(D'+\frac{D}{\rho}\right)
=\frac{2A}{\lambda}+\frac{3C}{\lambda}\rho^2\;,
\ee
in which case $H'=\lambda E$ implies $C=A\lambda^2/6$, which indeed matches the solution with 
$I_1$, 
\be
D=E\simeq A\rho\left(1+\frac{\lambda^2\rho^2}{6}\right).
\ee

This solution is indeed a solution without source, since again integrating (\ref{deltasource}) over a small disk as
before, we now find 
\be
A2\pi \epsilon^2=2A\pi \epsilon^2+C\Rightarrow C=0.
\ee

At $\rho\rightarrow \infty$, we also have two possible behaviours: the divergent one, to be excluded 
on physical grounds,
\be
E=I_1(\lambda\rho)\simeq \frac{e^{\lambda \rho}}{\sqrt{2\pi \lambda\rho}}\;,\;\;\
B=I_0(\lambda\rho)\simeq E\;,
\ee
and the good one, 
\be
E=K_1(\lambda \rho)\simeq e^{-\lambda\rho}\sqrt{\frac{\pi}{2\lambda \rho}}\;,\;\;\;
B=-K_0(\lambda \rho)\simeq E.
\ee

Note that at these large distances, the CS term dominates over the Maxwell one, hence the exponential 
behaviour (unlike the Maxwell behaviour, $E\simeq 1/\rho$).

Also note that, since the differential equation is linear, we have two solutions with general coefficients, 
but in the nonlinear case to be studied later, we {\em can} have uniquely fixed solutions (or not, 
depending on the nonlinear modification, as we will see).

We would like to determine for this solution the  charge, energy, momentum, angular momentum
and mean radius. The charge is the integral of the divergence of the electric field (in this Maxwell 
case there is no difference between $\vec{D}$ and $\vec{E}$). Ignoring for the moment the source charge at the 
origin, of value $C/g^2=2\pi \tilde b/(g^2\lambda)$, and integrating only until a small radius $\epsilon$ (since as we will 
see, the energy is divergent anyway, but both problems will be cured by going to the BI theory),\footnote{If we nevertheless 
include the charge at the origin, so including $r=0$ in our integration region, we obtain twice the charge, 
and so we find $J/Q=1/4$, i.e., if we fix $\tilde b$ such that $Q=N$, then we find $J=N/4$. But $r=0$ doesn't contribute to the 
charge in the correct BI case, so we will ignore it.} we obtain 
\be
Q = \frac{1}{g^2}\int_\epsilon d^2 x \nabla \cdot \vec E = \frac{\lambda}{g^2}\int_{S_\epsilon} d^2 x B = 2\pi \frac{
\tilde b}{ g^2\lambda} \int_{\epsilon\rightarrow 0}^\infty dz z K_0[z]= 2\pi \frac{\tilde b}{g^2\lambda}\;,
\ee
where  we have used $\int_0^\infty dz z K_0[z] =1$.

If we choose the constant to be $\tilde b= {\lambda}^2$, we get that 
\be
Q = N\;,
\ee
as we want.

For a radial electric field $E$, the components of the momentum $P_x$ and $P_y$ 
(given by the Poynting vector $\vec{\cal P}$) vanish.
 
 The angular momentum  $J$ is given by (the 4-dimensional $\vec{J}=\int \vec{r}\times \vec{P}$, with $\vec{P}=
 \vec{E}\times \vec{H}$ the Poynting vector becomes in 3 dimensions $J=\int d^2 x\epsilon^{ij}x_i P_j$, with 
 $P^i =\epsilon^{ij}E_j B/g^2$, and $x_i E_i=\rho E_\rho=\rho E$, so $J=\int d^2 x \rho EB/g^2$) 
\be
J= \frac{1}{g^2}\int d^2 x \rho E B=  \frac{2\pi}{g^2}  \frac{\tilde b^2 }{\lambda^3}\int_0^\infty dz z^2 K_0[z] K_1[z]
= \frac{2\pi}{g^2}   \frac{\tilde b^2 }{\lambda^3} \frac12 \;,
\ee
where we have used $\int_0^\infty dz z^2 K_0[z]  K_1[z]=\frac12$.
Upon  substituting the value of the constant $\tilde b$ chosen above, we get
\be
J=  \frac{2\pi}{g^2}   \frac{\tilde b^2 }{\lambda^3} \frac12  = \frac{2\pi}{g^2}   \frac{\lambda^4 }{\lambda^3} \frac12 =\frac{N}{2}.
\ee

Then the mean radius of the object  is given by
\be
\bar \rho=\frac{\frac{\lambda}{g^2}\int d^2 x \rho B }{\frac{\lambda}{g^2}\int d^2 x B }
=\frac{1}{\lambda}
\frac{\int_0^\infty dz z^2 K_0[z]}{\int_0^\infty dz z K_0[z]}=\frac{\pi}{2}\frac{1}{\lambda}\;,
\ee
so we see that in units of $\lambda$, which is the only parameter appearing in the equations of motion 
(\ref{eqom}) (note that, in this classical case we are considering, the equations of motion are the 
relevant object), the mean radius is independent of $N$.

Even though the object described by this static solution does 
not relate to the usual flavor degrees of freedom in the Sakai-Sugimoto-Witten (SSW) model 
\cite{Sakai:2004cn,Sakai:2005yt,Witten:1998zw}, it does
 admit properties similar to  what is expected in the large $N$ from the novel type of 
 baryon, namely it has $Q=N, J=N/2$ and $\bar\rho$ is independent of $N$. 
 
However, for it to represent a baryon as a soliton, ignoring the delta function source for a while, 
we still need to check the energy of the object. 
Calculating the energy (note that the CS term does not contribute to the Hamiltonian, hence to the 
energy, so the energy is the same as in the pure Maxwell case),
\be
{\cal E}= \frac{1}{2 g^2} \int d^2 x (E^2 + B^2)\;,
\ee
we obtain a divergence  of the integral near $\rho=0$, 
$\int_0^\infty E^2 \sim \int_0^\infty dz z K_1[z] K_1[z]$. However, the magnetic part of the energy is 
finite, since $\int_0^\infty dz z K_0[z] K_0[z]= \frac12$.

For future use, note the general formulas
\bea
\int_0^\infty x^\mu dx K_\nu (ax)&=&2^{\mu-1}a^{-\mu-1}\Gamma\left(\frac{1+\mu+\nu}{2}\right)
\Gamma\left(\frac{1+\mu-\nu}{2}\right)\cr
\int_0^\infty x^{-\lambda}dx K_\mu (ax)K_\nu (bx)&=&\frac{2^{-2-\lambda}a^{-\nu +\lambda-1}
b^\nu}{\Gamma(1-\lambda)}\Gamma\left(\frac{1-\lambda+\mu+\nu}{2}\right)\Gamma\left(\frac{1-
\lambda-\mu-\nu}{2}\right)\times\cr
&&\times \Gamma\left(\frac{1-\lambda+\mu-\nu}{2}\right)\times\cr
&&\times F\left(\frac{1-\lambda+\mu+\nu}{2},
\frac{1-\lambda-\mu+\nu}{2}; 1-\lambda;1-\frac{b^2}{a^2}\right).\cr
&&
\eea

\subsection{Regularization with a conducting circle around the origin}

The divergence of the energy, as well as the source,  come from the near $\rho=0$ region. To avoid them, we can 
consider a system with a conducting circle of radius $\rho_0$ around the origin, 
so that the electric and magnetic fields inside it vanish. 
Now all the integrals in the expressions for $Q, J,\bar \rho $ and ${\cal E}$ will be only
  between $\rho_0$ and infinity. 
  
If we take for this case that the constant is  $\tilde b=\frac{\lambda^2}{\hat Q}$, where 
$\hat Q =(\rho_0\lambda)K_1[(\rho_0\lambda)]$, which ensures that we still have $Q=N$, 
we get for the angular momentum 
\be
J=  \frac{2\pi}{g^2}   \frac{\tilde b^2 }{\lambda^3}\int_{\rho_0}^\infty dz z^2 K_0[z] K_1[z]  
= \frac{2\pi}{g^2}   \frac{\lambda^4 }{\hat Q^2\lambda^3}\frac12 \left[(\rho_0\lambda)K_1[(\rho_0\lambda)\right]^2   =\frac{N}{2}.
\ee

Thus, even for this regularized set-up, the ratio $\frac{J}{Q}= \frac12$ is still maintained. 

The  finite energy in this case is given by
\be
{\cal E}= \frac{1}{2 g^2} \int d^2 x ( E^2 + B^2)= \lambda N {\cal E}_0\;,
\ee
where the dimensionless quantity ${\cal E}_0$  is given by 
\be
{\cal E}_0= \frac{K_0[\rho_0 \lambda]}{2  (\rho_0\lambda) K_0[\rho_0 \lambda]}.
\ee

 The mean radius is now 
\be
\bar \rho=\frac{\frac{\lambda}{g^2}\int d^2 x \rho B }{\frac{\lambda}{g^2}\int d^2 x B }
=\frac{1}{\lambda}
\frac{\int_{\rho_0}^\infty dz z^2 K_0[z]}{\int_{\rho_0}^\infty dz z K_0[z]}
=\frac{\pi}{2}\frac{\hat \rho}{\lambda}\;,
\ee
where 
\be
\hat\rho = \frac{1}{6} \left(-3 \pi  \pmb{L}_2(\lambda\rho_0)+\frac{3 \pi 
   \left(\frac{1}{\lambda\rho_0}-\pmb{L}_1(\lambda\rho_0) K_2(\lambda
	\rho_0)\right)}{K_1(\lambda\rho_0)}+4 \lambda\rho_0\right)\;,
\ee 
and $\pmb{L}_1(z)$ and $\pmb{L}_2(z)$ are the modified Struve function of order one 
and two, respectively.

To conclude, in the ``regularized case" where the electric and magnetic fields vanish within 
a radius $\rho_0$ from  the origin, we can still get a solution that admits a charge $Q=N$, 
angular momentum $J=\frac{N}{2}$, while having now a finite energy, quantized
in terms of the scale $\lambda$ in the equations of motion,  
${\cal E}= \lambda N {\cal E}_0 $, and a mean radius that is $N$-independent, in terms of the scaling with 
$\bar \rho\sim \frac{1}{\lambda}$.

\subsection{Time-dependent solution}

We have found a static solution of the equations of motion (\ref{eqom}), but it had a divergent 
energy. Let us look now for  a time-dependent solution. In particular we would like to check 
whether there is  solution that incorporates a ``chiral mode", while keeping the same scalling 
of $Q,J$ and $\bar \rho$ with $N$.
We start with an ansatz that includes both a radial, as well as an azymuthal component 
of the electric field vector,
\be
\vec E = E_\rho \hat \rho + E_\theta \hat \theta\;, \qquad E_\rho= E_\rho(\rho)\;,   \qquad  E_\theta
= E_\theta(\rho)  \cos(\theta-wt). 
\ee
 
Since  $E_\theta$  now does  depend on theta, the divergence equation (the first equation in 
(\ref{eqom})) has  another term, so we also modify the ansatz for B in the form 
\be
B= B_\rho(\rho) + B_\theta(\rho,\theta) \;,
\ee
such that the additional equation that follows from the first equation of (\ref{eqom}) reads
\be\label{EBtheta}
\frac{1}{\rho} \pa_\theta E_\theta =\lambda  B_\theta \rightarrow \qquad  B_\theta=-\frac{1}{\lambda\rho} E_\theta(\rho) \sin(\theta-wt).
\ee

The second and third  equations now read
\bea
\pa_y B_\rho =\lambda E_\rho(\rho) \sin(\theta) & &  \pa_y B_\theta =\lambda E_\theta \cos(\theta)-\pa_t E_\theta \sin{\theta}\cr
\pa_x B_\rho = \lambda E_\rho(\rho)\cos(\theta)    & & \pa_x B_\theta =-\lambda E_\theta \sin(\theta)-\pa_t E_\theta \cos{\theta} \;,
\eea
from which it follows that 
\be\label{Brhotheta}
\pa_\rho B_\rho = \lambda E_\rho \qquad
 \pa_\rho B_\theta= -w E_\theta \sin(\theta-wt).
\ee

 Thus, it follow that  $E_\rho$ obeys the modified Bessel equation (\ref{BesselK}), namely
\be\label{BesselKtrho}
 \rho^2 E_\rho''  + \rho E_\rho ' - E_\rho( 1+ \lambda^2 \rho^2)= 0\;,
\ee
and hence we have
\be\label{sol}
E_\rho = \lambda^2 K_1[ \lambda\rho ] \qquad B_\rho = \lambda^3 K_0[ \lambda\rho]. 
\ee

As for $E_\theta$ and $B_\theta$, if we substitute the right-hand side of (\ref{EBtheta}) into 
the right-hand side of (\ref{Brhotheta}), we get  
\be
-\frac{\pa_\rho E_\theta}{\lambda\rho} + \frac{ E_\theta}{\lambda\rho^2}
=-w E_\theta \rightarrow \qquad  \rho\pa_\rho E_\theta -(w \lambda \rho^2 +1) E_\theta=0.
\ee

The solution of this equation is
\be
E_\theta=c \rho e^{\lambda w \rho^2}\qquad B_\theta= -\frac{c}{\lambda}e^{\lambda w \rho^2} 
\sin(\theta-wt).
\ee

The exponential growth of $E_\theta$ is surprising. Note that if one uses Euclidean instead 
of Lorentzian signature, this growth turns into a decay,
$e^{-\lambda w \rho^2}$.

Since when determining $Q$ we integrate over $\theta$, we get that if there is a natural cut-off 
 along $\rho$, $B_\theta$ does not contribute to $Q$, and thus we still have that 
\be
Q= N.
\ee

The angular momentum does not  involve $E_\theta$ and again the integral over $B_\theta$ vanishes,
so we also still get that 
\be
J= \frac{N}{2}.
\ee

We also get again that 
\be
\bar\rho = \frac{\pi}{2}\frac{1}{\lambda}.
\ee

\section{The 3-dimensional BI action plus CS term}

We want to find a finite energy soliton solution, so we must modify the action in the region where the divergence is 
situated, namely at $\rho\rightarrow 0$.

\subsection{Equations of motion and constitutive relations}

To obtain that, we replace the Maxwell term by a BI term. The main goal is to check whether the 
``soliton'' solution  (\ref{solCS}) is modified in the case of a BI action such that we have  
a finite energy, rather than a divergent one (as well as no delta function source). Consider then
\be
S_{\rm CS+BI}= \int d^3 x \left\{Rb^2\left[1-\sqrt{1+\frac{1}{2 g^2b^2}F_{\mu\nu}F^{\mu\nu}}\right] 
+ \frac{N}{2\pi} \epsilon^{\mu\nu\rho} A_\mu\pa_\nu A_\rho\right\}\;,
\ee
where $b$ has dimension 2, $R$ is a length scale and $g^2$ is dimensionless, so that $R/g^2$ is the 
previously defined $1/g^2$, now renamed $1/\tilde g^2$, that will continue to appear in $\lambda$.

The corresponding equations of motion are
\be
\pa_\nu\left(\frac {F^{\nu\mu}}{\sqrt{1+\frac{1}{2 g^2b^2}F_{\mu\nu}F^{\mu\nu}}}\right) + \lambda \epsilon^{\mu\nu\rho} F_{\nu\rho}=0.
\ee 

Explicitly, we have
\bea
\pa_1 \tilde D_1 + \pa_2 \tilde D_2 - \lambda B & = & 0 \cr
\pa_0 \tilde D_1 - \pa_2 \tilde H + \lambda E_2 & = &  0\cr
\pa_0 \tilde D_2 + \pa_1 \tilde H - \lambda E_1 & = & 0\;,
\eea
where
\bea
\tilde D_1= D_1g^2&=&\frac{E_1}{\sqrt{1-\frac{1}{g^2b^2}(E^2-B^2)}}\;, \qquad 
\tilde D_2= D_2 g^2=\frac{E_2}{\sqrt{1-\frac{1}{g^2b^2}(E^2-B^2)}}\;, \cr
\tilde H=Hg^2&=& \frac{B}{\sqrt{1-\frac{1}{g^2b^2}(E^2-B^2)}}\;,
\eea
and as usual (but referring only to the BI part, the CS term depends explicitly on $A_\mu$, 
so we cannot include it in the definition of $\vec{D},H$)
\be
\vec{D}=\frac{\d {\cal L}}{\d \vec{E}}\;,\;\;\; H=-\frac{\d {\cal L}}{\d B}.
\ee

Note that in 3 dimensions the magnetic field $B$ is a scalar, and so is $H$.

Looking for a static solution with only  a radial component of $\vec E$ denoted by $E$, 
the equations take the form
\bea
\frac {\tilde D}{\rho} + \tilde D' & = &  \lambda B \cr
\pa_y \tilde H & = &  \lambda \tilde E_y \cr
\pa_x \tilde H & = &  \lambda \tilde E_x\;, 
\eea
which imply the 2 regular differential equations for the radial fields,
\be
\tilde H'=\lambda E\;,\;\;\; \tilde D'+\frac{\tilde D}{\rho }=\lambda B.\label{eom}
\ee

A note on the BI action: When reducing the 4-dimensional BI action (\ref{LagDBI}) to 3 dimensions, 
two things happen: we are left with only $B=B_z$ and 
$E_1$ and $E_2$, so $B_1=B_2=0$, $E_z=0$, which also means that $\vec{E}\cdot \vec{B}=0$, 
hence $G=0$ now, and the second is that we integrate over $z$, giving a factor $R$ in front, with 
dimensions of length.  We also have introduced $1/g^2$ in front of $F$ in the action. 

Then, the constitutive relations become now (after absorbing the factor of $R$ in $\tilde D$ and $\tilde H$)
\bea
H(\vec{E},B)&=&\frac{1}{g^2}\frac{B}{\sqrt{1+\frac{B^2-\vec{E}^2}{g^2b^2}}}=\frac{\tilde H}{g^2}\cr
\vec{D}(\vec{E},H)&=&\frac{1}{g^2}\frac{\vec{E}}{\sqrt{1+\frac{B^2-\vec{E}^2}{g^2b^2}}}.\label{HD}
\eea

It would seem that we could simply use the above constitutive relations in (\ref{eom}), but that would 
be more difficult. It is clear that the better form is in terms of $\vec{D}, B$ and $\vec{E}(\vec{D},B)$
and $H(\vec{D},B)$, which are the natural variables in the Hamiltonian formalism. 

The Hamiltonian, as the Legendre transform of the Lagrangian, which in 4 dimensions was (\ref{HamLeg}), 
becomes in 3 dimensions
\be
{\cal H}=Rb^2\left[\frac{1+\frac{\vec{B}^2}{g^2b^2}}{\sqrt{1+\frac{B^2-\vec{E}^2}{g^2b^2}}}-1\right]\;,
\ee
but it needs to be re-expressed in terms of $\vec{D},B$, where $\vec{D}$ is now in (\ref{HD}).

Reducing to 3 dimensions the correct form of the Hamiltonian (\ref{HamCorrect}), 
in terms of $\vec{D}, \vec{B}$, 
we obtain
\be\label{HDB}
{\cal H}(b;\vec{D},\vec{B})=Rb^2\left[\sqrt{1+\frac{g^2\vec{D}^2+B^2/g^2}{b^2}
+\frac{\vec{D}^2B^2}{b^4}}-1\right].
\ee

Then 
\be
\vec{E}(\vec{D},B)=\frac{\d {\cal H}}{\d \vec{D}}=g^2\vec{D}\frac{1+B^2/(g^2b^2)}
{\sqrt{1+\frac{g^2\vec{D}^2+B^2/g^2}{b^2}
+\frac{\vec{D}^2B^2}{b^4}}}.
\ee

Moreover, since we can check that 
\be
\frac{\vec{E}^2}{g^2b^2}=\frac{g^2\vec{D}^2}{b^2}\frac{1+B^2/(g^2b^2)}
{1+g^2\vec{D}^2/b^2}\;,
\ee
then 
\be
\tilde H(\vec{D},B)=\frac{B}{\sqrt{1+\frac{B^2}{g^2b^2}-\frac{\vec{E}^2}{g^2b^2}}}
=B\sqrt{\frac{1+g^2\vec{D}^2/b^2}{1+B^2/(g^2b^2)}}.
\ee

Then we want to solve the equations of motion (\ref{eom}), with constitutive relations 
\bea
\vec{E}(\vec{D},B)&=&\frac{\d {\cal H}}{\d \vec{D}}=g^2\vec{D}\frac{1+B^2/(g^2b^2)}
{\sqrt{1+\frac{g^2\vec{D}^2+B^2/g^2}{b^2}
+\frac{\vec{D}^2B^2}{b^4}}}\cr
\tilde H(\vec{D},B)&=&B\sqrt{\frac{1+g^2\vec{D}^2/b^2}{1+B^2/(g^2b^2)}}.\label{constrel})
\eea

\subsection{The analysis of possible solutions}

These are 4 equations in $z=\lambda\rho$ with 4 unknowns, so they will admit a solution. 

However, the solution is hard to obtain. We will focus on the solution near $\rho=0$. We have 
shown that the expansion of the exact solution in the Maxwell case near $\rho=0$ can also 
be obtained as follows: we propose an {\em ansatz} for one of the fields (there $E$), and then find the 
other fields from one of the equations of motion, and the constitutive relations, and finally check if the 
remaining equation is satisfied. 

In this case, specifically we find it easier to write an ansatz for $\tilde D(\rho)$, then find $B$ from 
$\tilde D'+\tilde D/\rho=\lambda B$, then $E$ and $H$ from the constitutive relations, and finally 
check if the equation $\tilde H'=\lambda E $ is satisfied. 

Since there are only a small number of possible behaviours near $\rho=0$, once we find one that 
works, it is the correct one. 

As in the Maxwell case, we can have, near $\rho=0$, the solution that was excluded before, since it 
blew up at infinity, with $D=A\rho+C\rho^3$. For the moment we will ignore it, though it will turn out to 
be the only possibility in the end. 

First, an observation: for $D\rightarrow \infty$ and $B\rightarrow \infty$, the 
constitutive relations (\ref{constrel}) give 
\be
E\simeq B\;,\;\;\; \tilde H\simeq \tilde D\;,
\ee
which is the opposite of the small field result, for $D\rightarrow 0$, $B\rightarrow 0$, 
\be
E\simeq \tilde D\;,\;\;\; \tilde H\simeq B.
\ee

We consider the following possibilities:

\begin{itemize}

\item 1. We first try $\tilde D$ diverging as a power law, $\tilde D=A/\rho^\alpha$, $\alpha \neq 1$ 
and $\alpha> 0$.

Then we get 
\be
B=(1-\a)\frac{A}{\lambda \rho^{1+\a}}\;,\;\; \tilde H = \frac{A}{\rho^\a}sgn(\lambda)sgn(1-\a)\;,\;\;
E=|B|.
\ee

On the other hand, from the equation of motion, we have 
\be
\lambda E=\tilde H'=-\a\frac{A}{\rho^{1+\a}}sgn(1-\a).
\ee

We see that we have matching with the previous only if $\a\rightarrow \infty$. This actually means 
$\tilde D=Ae^{\frac{\b}{\rho^\gamma}}$, and we will comment on this later on, but for now, 
we will continue to try other cases.

\item 2. We can also have $\tilde D=A\ln \rho$, giving 
\be
B\simeq\frac{A}{\lambda}\frac{\ln \rho}{\rho}\;,\;\;\;
E=\frac{A}{|\lambda|}\frac{\ln \rho}{\rho}\;,\;\;\;
\tilde H=A \ln \rho sgn(\lambda).
\ee

But on the other hand, from the equation of motion, we get 
\be
\lambda E=\tilde H'=\frac{A}{\rho}sgn(\lambda)\;,
\ee
so it doesn't match. This is not a good solution.

\item 3. More generally, $\tilde D=A\ln^\a \rho$, gives
\be
B\simeq \frac{A}{\lambda}\frac{\ln^\a \rho}{\rho}\;,\;\;\;
\tilde H =A\ln^{\a}\rho sgn(\lambda)\;,\;\;
E=A\frac{\ln^\a\rho}{\rho}sgn(\lambda)\;,
\ee
but from the equations of motion, 
\be
\lambda E=\tilde H'=\a A \frac{\ln ^{\a-1}\rho}{\rho}sgn(\lambda)\;,
\ee
so this also doesn't match.

\item 4. We next try $\tilde D=A+K\rho^\a$, $\a>0$, giving 
\be
B\simeq \frac{A}{\lambda \rho}\;,\;\;
\tilde H =\frac{K}{\lambda \rho}\frac{A^2}{g^2b^2+A^2}\;,\;\;
E=\frac{A}{|\lambda|\sqrt{A^2+g^2b^2}}\frac{1}{\rho}\;,
\ee
and from the equation of motion
\be
\lambda E=\tilde H'<1/\rho\;,
\ee
so also doesn't match.

\item Similarly, we have also tried:
5. $\tilde D=A\rho^\a \ln \rho$, 6. $\tilde D=A+K\rho \ln \rho$, 7. $\tilde D=\tilde K/\ln \rho$, 
8. $\tilde D=A+\tilde K \rho^\a/\ln \rho$, 9. $\tilde D=A/\rho +C\ln \rho$, 
10. $\tilde D=A+\tilde K/\ln \rho$, 11. $\tilde D=A\rho^\a$, $\a>0$ (both $\a>1$ and $0<\a<1$). 
None of these works. 

\end{itemize}

This is good, since we can either have a unique solution, or two solutions, as in the Maxwell case, 
so if we find another possibility besides the $D=A\rho+C\rho^3$ one, that must be it. 

As we said, we could try ($\a,\b>0$) 
\be
\tilde D=Ae^{\frac{\a}{\rho^\b}}=-|\tilde D|\;,\;\;
B=-\frac{\a\b A}{\lambda\rho^{\b+1}}e^{\frac{\a}{\rho^\b}}=|B|\;
\ee
with $E\simeq B$ and $D\simeq H$. 

Note that now the Hamiltonian is 
\be
{\cal H}=Rb^2\left[\sqrt{\frac{\tilde D^2+B^2}{g^2b^2}+\frac{\tilde D^2B^2}{g^4b^4}}-1\right]\;,
\ee
so in our case it is 
\be
{\cal H}\simeq R\frac{\tilde D |B|}{g^2}\simeq \frac{A^2|\a\b|}{g^2|\lambda|}\frac{e^{\frac{2\a}{\rho
^\b}}}{\rho^{\b+1}}\;,
\ee
which {\em would} give an even more divergent energy! But now, unlike the purely electric 
BIon solution, for which we had to have $E/b\leq 1$ because of the square root $\sqrt{1-\vec{E}^2
/b^2}$, in this case, this doesn't contradict anything, since we have $\sqrt{1+B^2/b^2-\vec{E}^2/b^2}$,
and $B>E$. 

However, note that {\em while the leading behaviour in $B, D$ is OK, the subleading one gives a 
contradiction}! 

Indeed, if we are more precise, when $D\rightarrow\infty, B\rightarrow \infty$, from the 
constitutive relations  (\ref{constrel}), we have 
\bea
H&\simeq & D\left[1+{\cal O}\left(\frac{1}{B^2,D^2}\right)\right]\cr
E&\simeq & B\left[1+{\cal O}\left(\frac{1}{B^2,D^2}\right)\right].\label{divbeh}
\eea

In our case, using the leading behaviour of $D$ and $B$, we find
\bea
H&\simeq &Ae^{\frac{\a}{\rho^\b}}\left[1+{\cal O}\left(e^{-2\frac{\a}{\rho^\b}}\right)\right]\;,\cr
E&\simeq &-\frac{\a\b A}{\lambda \rho^{\b+1}}\left[1+{\cal O}\left(e^{-2\frac{\a}{\rho^\b}}\right)\right]\;.
\eea

On the other hand, from the equations of motion, $D'+D/\rho=\lambda B$ and $H'=\lambda E$, 
these two should reduce to (almost) the same equation, and by comparing the difference between the 
two, we find we should have
\be
\frac{D}{\rho}=\frac{A}{\rho}e^{\frac{\a}{\rho^\b}}={\cal O}\left(e^{-\frac{\a}{\rho^\b}}\right)\;,
\ee
which is a contradiction! 

So, in fact, there is no diverging solution either!

In this case, the only solution that we still have is 
the (modified) small field behaviour from the Maxwell case, which we also saw that had no delta function source. 
This corresponds to 
$D=A\rho+C\rho^3$, and we could prove it as above.

However, for ease of analysis in the case of other nonlinear actions besides BI, we will 
show how to derive them using the $D(E,B)$ and $H(E,B)$ formulas. 
In this case, we must make ansatze for {\em both} $E$ and $B$, then use the constitutive relations 
$D(E,B)$ and $H(E,B)$ and then check {\em both} equations of motion, $D'+D/\rho=\lambda B$ and 
$H'=\lambda E$.

At $\rho\rightarrow 0$, we write 
\be
E=A\rho+C\rho^3\;,\;\;\; B=B_0+B_2\rho^2.
\ee

From the constitutive relations, we get 
\be
D=\frac{1}{\sqrt{1+B_0^2}}(A\rho+C\rho^3)\;,\;\;\;
H=\frac{1}{\sqrt{1+B_0^2}}(B_0+B_2\rho^2).
\ee

The equation of motion $D'+D/\rho=\lambda B$ fixes
\be
B_0=\frac{2A}{\lambda \sqrt{1+B_0^2}}\;,\;\;\;
B_2=\frac{3C}{\sqrt{1+B_0^2}}\;,
\ee
while the equation of motion $H'=\lambda E$ fixes
\be
B_2=\frac{\lambda A\sqrt{1+B_0^2}}{2}\;,
\ee
so that
\bea
\frac{C}{A}&=& \frac{\lambda^2}{6}(1+B_0^2)\cr
B_2&=& \frac{\lambda^2A\sqrt{1+B_0^2}}{2}\cr
B_0\sqrt{1+B_0^2}&=&\frac{2A}{\lambda}.
\eea

Thus the solution is defined completely in terms of the arbitrary constant $A$, like in the Maxwell case.

At $\rho\rightarrow \infty$, we still have the exponentially small solution, 
we can ignore the BI modification to the action, since it vanishes at large distances.

But also at $\rho\rightarrow \infty$ 
we don't have the diverging solution anymore, for the same reason as in the small $\rho$ case. 
From (\ref{divbeh}) at large $\rho$, we need to be able to neglect $D/\rho$ with respect to $D'$, 
in order for the two equations of motion $D'+D/\rho=\lambda B$ and $H'=\lambda E$ to give the same
one in leading order. That excludes a power law, and only leaves an exponential in leading order, 
\be
D\simeq A e^{\a \rho^\b}\;,\;\; B\simeq \frac{A\a\b}{\lambda}\rho^{\b-1}e^{\a\rho^\b}\;,
\ee
with $\a,\b>0$. But then the subleading order doesn't match, since we get 
\be
\frac{D}{\rho}\simeq \frac{A}{\rho}e^{\a\rho^\b}\simeq {\cal O}\left(e^{-\a\rho}\right)\;,
\ee
which is a contradiction.

But then, the only possibility left is that there is a {\em unique solution}, with {\em $\rho\rightarrow 0$ 
behaviour given by the modified $I_1$ Maxwell solution at zero and the modified $K_1$ Maxwell 
solution at infinity}.  This will have a finite energy, as we wanted. 
One could in principle find this solution through numerical analysis, but this is left for further work.

We call the solution defined in this subsection the {\em CSBIon}.

\subsection{Charge, energy and angular momentum of the soliton solution}

We revisit the calculation of $Q,J, {\cal E}$ in Maxwell+CS theory, with a view to understand it in 
the case of the BI+CS soliton. 

We first note that, in general, $\vec{\nabla}\cdot \vec{D}=\rho_f$, the {\em free} (not polarization)
charge, usually $q\delta^d(\vec{r})$. 
But we also have the general Maxwell equation $\vec{\nabla}\cdot \vec{D}=\lambda B$ in the presence 
of the CS term, with no delta function source, so really we still have 
\be
Q=\frac{1}{g^2}\int d^2z \vec{\nabla}\cdot \vec{D}=\frac{\lambda}{g^2}\int d^2x B.
\ee

Here $\lambda=g^2N/(2\pi)$, and $E$ and $B$ are both proportional to an arbitrary constant, 
called $\tilde b$. 
In the Maxwell case, we {\em chose} it to be $=\lambda^2$, so that the charge $Q=N$. Now, for the 
same reason, we will choose a slightly different value. Note that $\tilde b$ has dimension 2, but 
once this is taken out, $E$ and $B$ become dimensionless functions of the dimensionless variable 
$z=\lambda \rho$. Thus we write 
\be
B=\tilde b B(z)\;,\;\;\; E=\tilde b E(z).
\ee

Note that, for the BI case,
\be
\tilde{\vec{D}}=\frac{\vec{E}}{\sqrt{1-\frac{\vec{E}^2-B^2}{g^2b^2}}}\;,
\tilde H=\frac{B}{\sqrt{1-\frac{\vec{E}^2-B^2}{g^2b^2}}}\;,
\ee
which means that also 
\be
\tilde{D}=\tilde b \tilde D(z)\;,\;\;
\tilde H=\tilde b \tilde H(z)\;,
\ee
and similarly for the case of the new relativistic modification in Appendix B.

Then 
\be
Q=\frac{2\pi \tilde b}{g^2}\int_0^\infty dz\; zB(z)\;,
\ee
where the integral is a dimensionless number, so we can now choose instead 
\be
\tilde b=\lambda^2\int_0^\infty dz\; zB(z)\Rightarrow Q=N.
\ee

The Poynting vector, giving the momentum density of the electromagnetic wave, is in 4 dimensions
\be
\vec{\cal P}=\vec{E}\times \vec{H}\;,
\ee
which in 3 dimensions becomes
\be
{\cal P}^i=\epsilon^{ij}E^j H\;,
\ee
and therefore the angular momentum is 
\be
J=\frac{1}{g^2}\int d^2x \rho E H=\frac{2\pi\tilde b^2}{g^2\lambda^3}\int_0^\infty dz\; z^2 E(z)H(z).
\ee

But with the above choice of $\tilde b$, we obtain 
\be
J=N\frac{\int_0^\infty dz\; z^2E(z)H(z)}{\left[\int_0^\infty dz\; z B(z)\right]^2}.
\ee

Unfortunately, without a full solution, we cannot calculate the coefficient of $N$ in the above. 

Because of the scaling of the fields with $\tilde b$ and $g$, and the form of the Hamiltonian ${\cal H}$, 
expanded in powers of the fields, we can write, in the Maxwell as well as in the BI (and new relativistic)
cases, 
\be
{\cal H}=\frac{1}{g^2}\tilde b^2 {\cal H}(z)\;,
\ee
 so that the (finite) energy is now
\be
{\cal E}=\frac{2\pi \tilde b^2}{g^2\lambda^2}\int_0^\infty dz\; z {\cal H}(z).
\ee

With the choice of $\tilde b$, we have now 
\be
{\cal E}=N\lambda\frac{\int_0^\infty dz\; z {\cal H}(z)}{\left[\int_0^\infty dz \; z B(z)\right]^2}.
\ee

Since $\lambda$ has dimension 1 and is the only dimensional constant appearing in the 
equations of motion, we can consider it as the scale of the energy although, strictly 
speaking, from the point of view of the action, where we have separately the dimension 1 constant
$g^2$ and $N$, $\lambda$ is quantized in units of $N$ as well, so the energy would be proportional to 
$N^2$, not $N$. 

The coefficients of $N$ in $J$ and $N\lambda$ in ${\cal E}$ can only be calculated numerically, or 
knowing the full solution.

\subsection{ModMax and ModMax precursor generalizations in 3 dimensions}

One can ask about the generality of the analysis in the Maxwell and BI cases. 

One could think that perhaps the new ModMax theory of Bandos et al. \cite{Bandos:2020jsw}, 
an extension of 
Maxwell with a dimensionless parameter $\gamma$, could also be of help 
in solving the singularity at $\rho=0$. 
We could extend the Maxwell term to the ModMax term, and we will do that
soon, but for the moment consider the more general precursor theory to ModMax, which is the 
theory that generalizes BI with the introduction of the same parameter $\gamma$, with Hamiltonian 
(see the Lagrangian in \cite{Nastase:2022fzx})
\be
{\cal H}^{(4d)}_{\rm BI-gen.}(\vec{D},\vec{B})
=\sqrt{T^2+2T \left(s\cosh \gamma -\sinh \gamma \sqrt{s^2-p^2}\right)+p^2}-T\;,
\ee
where 
\be
s=\frac{\vec{D}^2+\vec{B}^2}{2}\;,\;\;\; p =\sqrt{\vec{D}^2\vec{B}^2-(\vec{D}\cdot \vec{B})^2}.
\ee

Reducing to 3 dimensions, $\vec{B}$ becomes $B$, so we get
\be
s=\frac{\vec{D}^2+B^2}{2}\;,\;\;\; p =B|\vec{D}|\;,
\ee
and so
\be
\sqrt{s^2-p^2}=\sqrt{\left(\frac{\vec{D}^2+B^2}{2}\right)^2-\vec{D}^2B^2}
=\frac{|\vec{D}^2-B^2|}{2}.
\ee

Also introducing $g^2$, the 3 dimensional Hamiltonian is 
\be
{\cal H}^{(3d)}_{\rm BI-gen}(\vec{D},B)=R\sqrt{T^2+\frac{2T}{g^2}\left(\cosh \gamma \frac{\tilde
{\vec{D}}^2+B^2}{2}-\sinh \gamma \frac{|\tilde{\vec{D}}-B^2|}{2}\right)+\frac{\tilde{\vec{D}}^2
B^2}{g^4}}-RT.
\ee

To this, one must, of course, add the CS Hamiltonian, but that vanishes, since the CS Lagrangian 
is linear in $\dot{\vec{A}}$ (it has the term $\dot A_1 A_2-\dot A_2 A_1$), so we are safe. 

Then we define $\vec{E}$ and $H$ as usual, obtaining 
\bea
\vec{E}&=& \frac{\d {\cal H}}{\d \vec{D}}=\tilde{\vec{D}}\frac{T\left[\cosh \gamma B
-\sinh \gamma {\rm sgn}(\tilde{\vec{D}}^2-B^2)\right]+B^2/g^2}{
\sqrt{T^2+\frac{2T}{g^2}\left(\cosh \gamma \frac{\tilde
{\vec{D}}^2+B^2}{2}-\sinh \gamma \frac{|\tilde{\vec{D}}-B^2|}{2}\right)+\frac{\tilde{\vec{D}}^2
B^2}{g^4}}}\cr
\tilde{H}&=&\frac{\d {\cal H}}{\d B}= B \frac{T\left[\cosh \gamma B
-\sinh \gamma {\rm sgn}(\tilde{\vec{D}}^2-B^2)\right]+\tilde{\vec{D}}^2/g^2}{
\sqrt{T^2+\frac{2T}{g^2}\left(\cosh \gamma \frac{\tilde
{\vec{D}}^2+B^2}{2}-\sinh \gamma \frac{|\tilde{\vec{D}}-B^2|}{2}\right)+\frac{\tilde{\vec{D}}^2
B^2}{g^4}}}.\label{constit}
\eea

The ModMax part of the Lagrangian is 
\be
{\cal L}(\vec{E},\vec{B})=T\left[1-\sqrt{1-\frac{B^2-\vec{E}^2}{g^2T}\cosh \gamma
-\sinh \gamma \frac{|B^2-\vec{E}^2|}{T^2}}\right]\;,
\ee
to which now we must add the CS term.

The equations of motion are, as in the BI case, 
\be
\frac{D}{\rho}+D'=\lambda B\;,\;\;\;
\tilde{H}'=\lambda E.
\ee

From the constitutive relations (\ref{constit}), we see that as $B,\tilde D\rightarrow\infty$, we obtain 
again 
\be
E\rightarrow B\;,\;\; \tilde H=\tilde D\;,
\ee
as in the BI case (the opposite of the small field results).

We also obtain that in the ModMax limit $T\rightarrow \infty$, the constitutive relations (\ref{constit}) 
become
\bea
\vec{E}&=& \tilde D\left[\cosh\gamma -\sinh\gamma {\rm sgn}(\tilde{\vec{D}}^2-B^2)\right]\cr
\tilde D&=& B \left[\cosh \gamma -\sinh\gamma {\rm sgn}(\tilde{\vec{D}}^2-B^2)\right].
\eea

This means that, up to a numerical factor, we are back to the constitutive relations of the Maxwell theory, 
so the same analysis as there follows. 

Instead, we may hope that the precursor to the ModMax has a better chance of avoiding the 
singularity, so we repeat the same analysis. But since we have $E\simeq B$ and $H\simeq D$ at 
large $D$ and $B$, we have the same analysis as in the BI case: the equations of motion in terms
of $E,B, D,H$ are the same, and for diverging $D,B$ the same constitutive relations, so again 
we take $D=A/\rho^\a$ and (since then $D$ and $B$ are large) find matching only for $\a\rightarrow
\infty$. 

Moreover, then explicitly again we can take $D=Ae^{\a/\rho^\b}$ and obtain matching, but 
only for the leading order, the subleading one doesn't work.  So in this case again we have 
a solution interpolating between the modified Maxwell $I_1$ solution at $\rho=0$ and the modified 
Maxwell $K_1$ solution at $\rho\rightarrow \infty$.  This again gives a finite energy.

In order to find the generality of the solution to the diverging energy problem in nonlinear theories
of electromagnetism, we consider other nonlinear modifications in the Appendices.


\section{Conclusions and discussion}

In this paper we have defined a finite energy solution to 3-dimensional BI+CS electromagnetism 
(abelian gauge theory), which we called a CSBIon. The solution for a level $N$ CS term has charge $N$, 
radius that is $N$-independent, and angular momentum and (finite) energy proportional to $N$, 
which means the solution represents a soliton.

The CS+BI theory was understood heuristically in string theory as a D6-brane wrapping an $S^4$ in a D4-brane
background, giving the CS+BI theory on the common D2-brane worldvolume. 

The CS term is crucial in many condensed matter applications, since it dominates at low energies over the Maxwell 
term. But it was crucial for the finiteness of the soliton that we had BI electromagnetism, not Maxwell. However, we 
can understand the BI modification as a type of regularization. In fact, since the BI scale is related to the string 
scale in string theory, the regularization appears because of string theory. 

The list of open questions related to the CSBIon  include in particular the following ones:
\begin{itemize}
\item
Deriving  explicit, probably numerical, solutions of the equations of motion of the BI+CS theory.
\item
In this paper we have analyzed the pure gauge theory. An obvious question is to consider the coupling 
of the BI+CS theory to scalar and fermion fields. It will be interesting to explore the interactions between 
the CSBIon and the matter fields.
\item
A natural generalization of the model discussed here is in the form of a non-abelian BI+CS theory.
\item
The action of the BI+CS emerges as the low energy effective action associated with D-branes in various 
string backgrounds. In these cases one needs to study the system in a curved background with possibly additional fields.
\item
Probably the most interesting question regarding the CSBIon is to find realizations of it in the context of condensed matter systems.
\item
In this paper we have analyzed the system classically. An obvious question is how to quantize it. 
\end{itemize}

\section*{Acknowledgements}
JS would like to thank O. Aharony, F. Bigazzi, A. Cotrone and Z. Komargodski for discussions regarding a related project. 
We would like to thank Z. Komargodski for his remarks on the manuscript.

The work of HN is supported in part by  CNPq grant 301491/2019-4 and FAPESP grants 2019/21281-4 
and 2019/13231-7. HN would also like to thank the ICTP-SAIFR for their support through FAPESP grant 2016/01343-7. 
The work of J.S was supported  by a center of excellence of the Israel Science Foundation (grant number 2289/18).
 
\appendix

\section{Nonrelativistic BI-type model}

We saw that the problem with the Maxwell modification to BI, 
and its ModMax precursor generalization, is that we have 
$\sqrt{1+B^2-\vec{E}^2}$ in the Lagrangian, which {\em in principle} 
allows for the solution where $E\simeq B\rightarrow
\infty$, unlike the case of the original BI purely electric solution, where effectively we had 
$\sqrt{1-\vec{E}^2}$, so $|\vec{E}|$ was bounded by 1 (and in fact it reached this value at the 
core of the BIon). That is why, although in fact we find that the diverging solutions are not allowed by 
the equation of motion, the finite energy solutions that we find have are not like in the case of the 
BIon, namely they do not go to a fixed, maximal, solution at $\rho=0$, but rather they go to a 
solution depending on an arbitrary constant.

Then, in order to have a solution with naturally bounded $|\vec{E}|$, 
as well as naturally bounded $B$, so with a more intuitive finite energy solution,
it suffices to reverse the sign of $B^2$ in the BI-type Lagrangian. To preserve the Maxwell Lagrangian 
at small fields, we also add a $B^2$ term, obtaining
\be
S_{CS+BI}^{\rm NR}
= \int d^3 x \left\{Rb^2\left[1-\sqrt{1-\frac{1}{g^2b^2}\left(B^2+\vec{E}^2\right)}-\frac{B^2}{g^2
b^2}\right] + \frac{N}{2\pi} \epsilon^{\mu\nu\rho} A_\mu\pa_\nu A_\rho\right\}\;.
\ee

Then we find
\bea
\tilde{\vec{D}}&=&g^2\frac{\d {\cal L}}{\d \vec{E}}=\frac{\vec{E}}{\sqrt{1-\frac{B^2+\vec{E}^2}{g^2
b^2}}}\cr
\tilde H&=& -g^2\frac{\d {\cal L}}{\d B}=2B -\frac{B}{\sqrt{1-\frac{B^2+\vec{E}^2}{g^2
b^2}}}.
\eea

The Hamiltonian is now
\be
{\cal H}=\vec{E}\vec{D}-{\cal L}=Rb^2\left[\frac{1-\frac{\vec{B}^2}{g^2b^2}}{\sqrt{1-\frac{B^2
+\vec{E}^2}{g^2b^2}}}-1+\frac{B^2}{g^2b^2}\right]\;,\label{corrham}
\ee
and as before, we find that we can rewrite it as 
\be
{\cal H}(b;\vec{D},\vec{B})=Rb^2\left[\sqrt{1+\frac{g^2\vec{D}^2-B^2/g^2}{b^2}
-\frac{\vec{D}^2B^2}{b^4}}-1+\frac{B^2}{g^2b^2}\right].
\ee
Then we have
\be
\vec{E}(\vec{D},B)=\frac{\d {\cal H}}{\d \vec{D}}=\tilde{\vec{D}}\frac{1-B^2/(g^2b^2)}
{\sqrt{1+\frac{\tilde{\vec{D}}^2-B^2}{g^2b^2}
-\frac{\tilde{\vec{D}}^2B^2}{g^4b^4}}}.
\ee

Moreover, since we can check that 
\be
\frac{\vec{E}^2}{g^2b^2}=\frac{\tilde{\vec{D}}^2}{g^2b^2}\frac{1-B^2/(g^2b^2)}{1+
\tilde{\vec{D}}^2/(g^2b^2)}
\;,
\ee
then 
\be
\tilde H(\vec{D},B)=2B-\frac{B}{\sqrt{1-\frac{B^2}{g^2b^2}-\frac{\vec{E}^2}{g^2b^2}}}
=2B-B\sqrt{\frac{1+\tilde{\vec{D}}^2/(g^2b^2)}{1-B^2/(g^2b^2)}}.
\ee

Then we want to solve the equations of motion (\ref{eom}), with constitutive relations 
\bea
\vec{E}(\vec{D},B)&=&\frac{\d {\cal H}}{\d \vec{D}}=\tilde{\vec{D}}\frac{1-B^2/(g^2b^2)}
{\sqrt{1+\frac{\tilde{\vec{D}}^2-B^2}{g^2b^2}
-\frac{\tilde{\vec{D}}^2B^2}{g^4b^4}}}\cr
\tilde H(\vec{D},B)&=&B\left[2-\sqrt{\frac{1+\tilde{\vec{D}}^2/(g^2b^2)}{1-B^2/(g^2b^2)}}\right].
\eea

Since we have the bound $|\vec{E}/(gb)|\leq 1$ and $|B/(gb)|\leq 1$ from the square root in the 
Lagrangian, if follows that $E$ and $B$ can at most be finite, but cannot be infinite. 

1. According to our recipe, we start with an ansatz for $D$. Assume first it is infinite, while $B$ must be
finite, as we said. Since $D'+D/\rho=\lambda B$, this is only possible if $D=A/\rho+C\rho+...$, which 
gives $B=2C/\lambda +...$ But then the constitutive relations give
\be
E=\tilde D\sqrt{1-\frac{B^2}{g^2b^2}}\;,\;\; \tilde H=2B-B\frac{|\tilde{\vec{D}}|/gb}{\sqrt{1-\frac{B^2
}{g^2b^2}}}.
\ee
Then 
\be
E\simeq \frac{A}{\rho}\sqrt{1-\frac{4C^2}{\lambda^2}}\;,\;\;\;
\tilde H \simeq -\frac{2C}{\lambda}\frac{A}{\rho\sqrt{1-\frac{4C^2}{\lambda^2}}}\;,
\ee
and we see that then $\tilde H\rightarrow \infty$ and moreover, $\tilde H'=E\rightarrow \infty$, which 
is not possible. So this possibility is out. 

From now on, we will consider $gb=1$ for simplicity (though it can be reinstated easily).

2. More generically, consider $E$ and $B$ finite, but $D$ non-infinite. Then 
\be
B=A+K\rho^\b\Rightarrow D=C\rho+K' \rho^{1+\b}.
\ee

But the constitutive relations then say 
\be
E\simeq D\sqrt{1-B^2}\sim C\rho\sqrt{1-A^2}\rightarrow 0\;,
\ee
so we get a contradiction. We could continue with $E\propto \rho$, and we will in fact see that this is
the solution, but for the moment we just say that $E$ cannot be finite if $B$ is finite.

3. We could have $E$ finite, but $B=K\rho^\a\rightarrow 0$, which would imply 
\be
D=C\rho^{1+\a}(1+K'\rho^\b)\;,
\ee
but then from the constitutive relations $H\simeq B\simeq K \rho^\a$ and $E\simeq D\simeq C\rho^{1+
\a}\rightarrow 0$, contradicting our assumption. 

4. We are left with the possibility that $E\rightarrow 0$ and $B$ finite. Assume
\be
B=A+K'\rho^\b\;,
\ee
which means that 
\be
D=C\rho(1+K\rho^\b)\rightarrow 0\;,
\ee
which gives 
\be
B=\frac{1}{\lambda}\left(D'+\frac{D}{\rho}\right)=\frac{2C}{\lambda}+\frac{(2+\b)CK}{\lambda}\rho
^\b.
\ee

But then, from the constitutive relations, 
\be
H\simeq B \left[2-\frac{1}{\sqrt{1-B^2}}\right]\;,
\ee
yet we want at least $H=F+G\rho^2$, so $E\propto \rho\rightarrow 0$. This implies $\b=2$ (at least), and 
moreover we can calculate $H$. We have two possibilities:

a) $F=0$, so $H\propto \rho^2$. In that case, we obtain 
\be
A=\frac{2C}{\lambda}=\frac{\sqrt{3}}{2}\Rightarrow K'=\frac{4Ck}{\lambda}=\sqrt{3}K.
\ee
Then also 
\be
H\simeq \frac{\sqrt{3}}{2}\left[2-\frac{2}{\sqrt{1-4\sqrt{3}K'\rho^2}}\right]\simeq -6 K'\rho^2.
\ee

From the consititutive relations, we obtain 
\be
E\simeq D\sqrt{1-B^2}\simeq \frac{C\rho}{2}\;,
\ee
but from the last equation of motion, we get
\be
E=\frac{H'}{\lambda}=-12\frac{K'}{\lambda}\rho=-12\sqrt{3}\frac{K}{\lambda}\rho\;,
\ee
and equating the two, we get
\be
K=-\frac{\lambda^2}{96}.
\ee

Then finally
\be
E\simeq \frac{\sqrt{3}}{8}\lambda \rho\;,\;\;\; B\simeq \frac{\sqrt{3}}{2}\left[1-\frac{\lambda^2\rho^2}{
48}\right]\;,
\ee
which gives a finite energy density at zero from (\ref{corrham}), just like for the BIon.

b) The more general case is for $F\neq 0$, so 
\bea
H&= &B \left[2-\frac{\sqrt{1+D^2}}{\sqrt{1-B^2}}\right]\cr
&\simeq& A\left\{2-\frac{1}{\sqrt{1-A^2}}+\rho^2\left[\frac{K'}{A}\left(2-\frac{1}{\sqrt{1-A^2}}\right)
-\frac{1}{\sqrt{1-A^2}}\left(\frac{C^2}{2}+\frac{K'A}{1-A^2}\right)\right]\right\}\cr
&=& \frac{2C}{\lambda}\left\{2-\frac{1}{\sqrt{1-(2C/\lambda)^2}}+\rho^2\left[2K\left(2-\frac{1}{
\sqrt{1-(2C/\lambda)^2}}\right)\right.\right.\cr
&&\left.\left.-\frac{1}{\sqrt{1-(2C/\lambda)^2}}\left(\frac{C^2}{2}+\frac{2K (2C/\lambda)^2}{1-
(2C/\lambda)^2}\right)\right]\right\}.
\eea

But from the constitutive relations we have 
\be
E\simeq D\sqrt{1-B^2}=C\rho\sqrt{1-(2C/\lambda)^2}\;,
\ee
while from the equation of motion we have
\be
E=\frac{H'}{\lambda}=\frac{4C\rho}{\lambda^2}
\left[2K\left(2-\frac{1}{
\sqrt{1-(2C/\lambda)^2}}\right)
-\frac{1}{\sqrt{1-(2C/\lambda)^2}}\left(\frac{C^2}{2}+\frac{2K (2C/\lambda)^2}{1-
(2C/\lambda)^2}\right)\right].
\ee

Equating the two, we obtain 
\be
K=\frac{1}{2}\frac{\frac{C^2}{2\sqrt{1-(2C/\lambda)^2}}+\frac{\lambda^2}{4}\sqrt{1-(2C/\lambda)^2}}
{2-\frac{1}{\sqrt{1-(2C/\lambda)^2}}-\frac{(2C/\lambda)^2}{(1-(2C/\lambda)^2)^{3/2}}}.
\ee

Thus we have obtained $K=K(C)$, and we had previously obtained
\be
B\simeq \frac{2C}{\lambda}(1-2K\rho^2)\;,\;\;\; E\simeq C\rho \sqrt{1-(2C/\lambda)^2}\;,
\ee
so the solution has a free parameter $C$, bounded by $C\leq \lambda/2$. That is good, since we have
solutions at infinity that are also defined by a free parameter. This is also what happens for the 
BIon solution.

\section{Relativistic BI-type model}

We can also consider relativistic nonlinear electromagnetism Lagrangians, but we consider one 
that obtains a stronger bound on the fields than in the BI case. We take
\be
\frac{1}{R}{\cal L}=\frac{\vec{E}^2-B^2}{2g^2}\sqrt{1-\left(\frac{\vec{E}^2-B^2}{g^2b^2}\right)^2}.
\ee

This guarantees that at least $|\vec{E}^2-B^2|\leq g^2b^2$, unlike the BI case, where, if 
$B$ diverges faster than $E$, $B$ can diverge as much as possible, as well as having $B^2-\vec{E}^2$
diverge as well. 

But we still have the problem that $\vec{E}^2$ and $B^2$ could diverge, as long as their difference 
doesn't, which would still give a divergent energy. 

First, we calculate the constitutive relations
\bea
\tilde{\vec{ D}}&=& 
g^2\frac{\d {\cal L}}{\d \vec{E}}=\frac{\vec{E}}{\sqrt{1-\left(\frac{\vec{E}^2-B^2}{g^2b^2}
\right)^2}}\cr
\tilde H &=& -g^2\frac{\d {\cal L}}{\d B}=\frac{B}{\sqrt{1-\left(\frac{\vec{E}^2-B^2}{g^2b^2}
\right)^2}}.\label{constDH}
\eea

Then the Hamiltonian is 
\be
{\cal H}=\vec{E}\cdot \vec{D}-{\cal L}=\frac{Rb^2}{\sqrt{1-\left(\frac{\vec{E}^2-B^2}{g^2b^2}
\right)^2}}\left[\frac{\vec{E}^2}{g^2b^2}+\frac{B^2}{g^2b^2}+\left(\frac{\vec{E}^2-B^2}{2g^2
b^2}\right)^3\right]\;,
\ee
just that now we haven't been able to rewrite it in terms of $\tilde D,B$, and find from it $\vec{E}(\vec{D},
B)$ and $H(\vec{D},B)$ as in the BI case. 

As a result, it is more difficult to analyze the solution to the equations of motion. Before, we had to 
only write an ansatz for $D$, then derive $B$ from the equations of motion, then $E$ and $H$ from the 
constitutive relations, and finally check the remaining equation of motion $H'=\lambda E$. 

Now, we must write {\em two } ans\"{a}tze, for $E$ and $B$, derive $D$ and $H$ from the constitutive 
relations, and finally check {\em both} equations of motion. 

But, because of the form of the Lagrangian, now this procedure is more doable. 

Indeed, now, if $B$ or $E$ is infinite, so must the other one, and we must have $B\simeq E\rightarrow 
\infty$, with $(B^2-E^2)^2\leq 1$. 

Let us assume that this is the case, and then we must also have, for the subleading terms,
{\em first in the case of $|B^2-E^2|\simeq1$},
\be
|B^2-E^2|=1-A\rho^\a\;,
\ee
which gives, from the constitutive relations (\ref{constDH}),
\be
D\simeq A'\frac{E}{\rho^{\a/2}}\simeq H.
\ee

Even in the case of $|E^2-B^2|=C\leq 1$, we still obtain $D\simeq H> B\simeq E$ (otherwise we 
have $\gg$ instead of just $>$, but the effect is the same).

We then obtain a contradiction, since on the one hand we have obtained $D\simeq H> B \simeq E$, 
but then from the equations of motion we have $|D|/\rho<|D'|$ in order to be able to neglect the 
extra term $D/\rho$ and the two equations of motion to give the same thing, and on the other 
we have then $|D'|\simeq \lambda E<\lambda D$, which finally gives $|D|/\rho<\lambda |D|$, which 
is a contradiction. 

So we don't can't have diverging fields at $\rho=0$, just like in the BI case, and for a similar reason. 
But we also can't have diverging fields at $\rho=\infty$, now called $r$ to remember 
that it goes to infinity, just like in the BI case.

Indeed, again we need to be able to neglect $D/r$ with respect to $D'$, in order to obtain 
the same equation of motion for the two, $D'+D/r=\lambda B$ and $H'=\lambda E$, since 
$E\simeq B$, say with subleading terms in a Taylor expansion,
\be
B^2-E^2=1-\frac{A}{r}\;,
\ee
so from the constitutive relations (\ref{constDH}),
\be
D\simeq \frac{E}{\sqrt{\frac{2A}{r}}}\simeq H\simeq \frac{B}{\sqrt{\frac{2A}{r}}}.
\ee

But for a diverging power law, $D'\sim D/r$, so we must have an exponential instead, 
\be
B\simeq E\simeq Ce^{\a r^\b}\;,
\ee
with $\b>0$. Moreover, then the equations of motion reduce in leading order to 
\be
D'\simeq\frac{C\a\b}{\sqrt{2A}}r^{\b-1/2}e^{\a r^\b}\;,
\ee
and equating with $\lambda B$ gives $\b=1/2$ and 
\be
\frac{\a\b}{\sqrt{2A}}=\lambda\Rightarrow A=\frac{\a^2}{8\lambda^2}.
\ee

It would seem like we found a solution, but in fact the solution is not valid for the subleading 
terms, which give a contradiction. Indeed, from the subleading terms for the two equations of motion, 
we obtain
\bea
\frac{3}{2}\frac{C\a}{\sqrt{2A}}\frac{e^{\a \sqrt{r}}}{\sqrt{r}}&=&\delta(\lambda B)\cr
\frac{1}{2}\frac{C\a}{\sqrt{2}}\frac{e^{\a\sqrt{r}}}{\sqrt{r}}&=&\delta(\lambda E)\;,
\eea
which would give 
\be
B^2-E^2\sim \frac{e^{\a\sqrt{r}}}{\sqrt{r}}\rightarrow \infty\;,
\ee
contradicting our assumption. So the diverging solution is excluded also at infinity. 

On the other hand, as usual, at infinity the exponentially small solution, given by CS+Maxwell, is still 
OK, since we can neglect the correction to the Maxwell action. 

And at $\rho=0$, again (like in the BI case) we have just a modification of the $I_1$ solution of 
the Maxwell case. Indeed, with the ansatz 
\bea
E&\simeq & A\rho+C\rho^3\cr
B&\simeq & B_0+B_2\rho^2\;,
\eea
from the constitutive relations (\ref{constDH}), we find
\bea
D&\simeq & \frac{1}{\sqrt{1-B_0^4}}(A\rho+B\rho^3)\cr
H&\simeq & \frac{1}{\sqrt{1-B_0^4}}(B_0+B_2\rho^2).
\eea

Then the equation of motion $D'+D/\rho=\lambda B$ gives
\be
\frac{2A}{\sqrt{1-B_0^4}}+\frac{3C}{\sqrt{1-B_0^4}}\rho^2=\lambda(B_0+B_2\rho^2)\;,
\ee
fixing
\be
B_0=\frac{2A}{\lambda\sqrt{1-B_0^4}}\;,\;\; 
B_2=\frac{3C}{\lambda\sqrt{1-B_0^4}}\;,
\ee
with $B_0$ solving therefore the equation 
\be
B_0\sqrt{1-B_0^4}=\frac{2A}{\lambda}\;,
\ee
while the $H'=\lambda E$ equation gives
\be
\frac{C}{A}=\frac{\lambda^2}{6}(1-B_0^4)\;,
\ee
so that finally
\be
B_2=\frac{\lambda A\sqrt{1-B_0^4}}{2}\;,
\ee
so all the coefficients are written in terms of a single one, $A$, like in the BI case. 

We can also easily exclude the other potential cases at $\rho\rightarrow 0$ and 
$B\rightarrow$ constant (which implies $H\propto \rho$, plus maybe a constant from $H'=\lambda \rho$,
but that forces the square root in the action to be finite, which in turn means $D$
starts with a constant, but then the term $D/\rho$ in the equation of motion $\lambda B=D'+D/\rho$
implies a diverging term in $B$, contradiction), 
or $E\rightarrow $constant and $B\rightarrow 0$ (which again imples the $D/\rho$ term  for $\lambda B$
must diverge, giving a contradiction), as not solving the 
equations of motion, just as they were excluded in the BI case.

That means again, like in the BI case, that the non-diverging solution, modification of the $I_1$ 
solution in the Maxwell case at $\rho=0$, matches onto the non-diverging solution, modification of 
the $K_1$ solution at $\rho=\infty$, giving a finite energy solution.


\section{Attempts of finding an analytic solution} 

In this Appendix we try to see if we can {\em guess} a full solution of the equations of motion in the BI+CS case, 
based on the Maxwell+CS solutions, and what happens in 3+1 dimensions if we change the Maxwell theory into a 
BI theory.

We first note that the solutions of the BI+CS turn into the solutions of the Maxwell+CS in the asymptotic limit 
$\rho\lambda\rightarrow \infty$. Thus, an idea is to take an ansatz for the solutions of the BI+CS in the form of
\be
D(\rho\lambda)= K_1(\rho\lambda)f(\rho\lambda)\;,\qquad \lim_{\rho\lambda\rightarrow \infty}f(\rho\lambda)= 1
\ee
\be
B(\rho\lambda)= -K_0(\rho\lambda)(f(\rho\lambda)+g(\rho\lambda))\;,\qquad 
\lim_{\rho\lambda\rightarrow \infty} g(\rho\lambda)= 0.
\ee

Upon inserting this into (\ref{eom}) we find that 
\be
f'(\rho\lambda)=-\lambda \frac{K_0}{K_1}g(\rho\lambda).
\ee

With this ansatz for $D$ and $B$, we get that $E$ and $H$ take the form
\be
E(\rho\lambda)=f(r) K_1(r) \sqrt{\frac{K_0(r){}^2
   (f(r)+g(r))^2+1}{f(r)^2 K_1(r){}^2+1}}
	\ee
and 
\be
H = K_0(r) (f(r)+g(r)) \sqrt{\frac{f(r)^2
   K_1(r){}^2+1}{K_0(r){}^2 (f(r)+g(r))^2+1}}.
\ee

Pluging these expression into the eom (\ref{eom}) we find the following constraint equation on $f$ and $g$

\bea
&-&\frac{\sqrt{\frac{f(r)^2 K_1(r){}^2+1}{K_0(r){}^2
   (f(r)+g(r))^2+1}} \sqrt{\frac{K_0(r){}^2
   (f(r)+g(r))^2+1}{f(r)^2 K_1(r){}^2+1}}}{f(r)
   K_1(r) 
	\left(K_0(r){}^2
   (f(r)+g(r))^2+1\right){}^2}\times
\CR
&\times&
\frac{1}{r}\left(f(r) K_0(r) K_1(r){}^2 \left(-r g(r)
   f'(r)+K_0(r){}^2 (f(r)+g(r))^3 \left(f(r)-r
   f'(r)\right)\right.\right.\cr
   &&\left.\left.-2 r f(r) f'(r)-r f(r) g'(r)+f(r)
   g(r)+f(r)^2\right)\right)
\CR
&\times&\left(
	-K_0(r)
   \left(f'(r)+g'(r)\right)+f(r)^2 K_1(r){}^3
   (f(r)+g(r))\right.\cr
   &&\left.+K_1(r) (f(r)+g(r)) \left(f(r)^2
   K_0(r){}^2 \left(K_0(r){}^2
   (f(r)+g(r))^2+1\right)+1\right)\right)
\CR
&=& {f(r)
   K_1(r) \left(K_0(r){}^2
   (f(r)+g(r))^2+1\right){}^2}.
 \eea

\subsection{Attempt 1}

The simplest attempt is obviously to take
\be
f(\rho\lambda)=1\;, \qquad g(\rho\lambda)=0.
\ee

In fact this can be generalized, since 
the equation of motion (\ref{eom}) that relates $\tilde D$ and $B$,
\be
\tilde D'+\frac{\tilde D}{\rho }=\lambda B\;,
\ee
has a solution of the form
\be
\tilde D = \tilde b(c K_1[ \lambda\rho ]+ d I_1[ \lambda\rho ]\qquad B =  \tilde b(c K_0[ \lambda\rho ]-d I_0[ \lambda\rho ]).
\ee

\begin{figure}[t!]
\begin{tabular} {c}
\includegraphics[width=6cm]{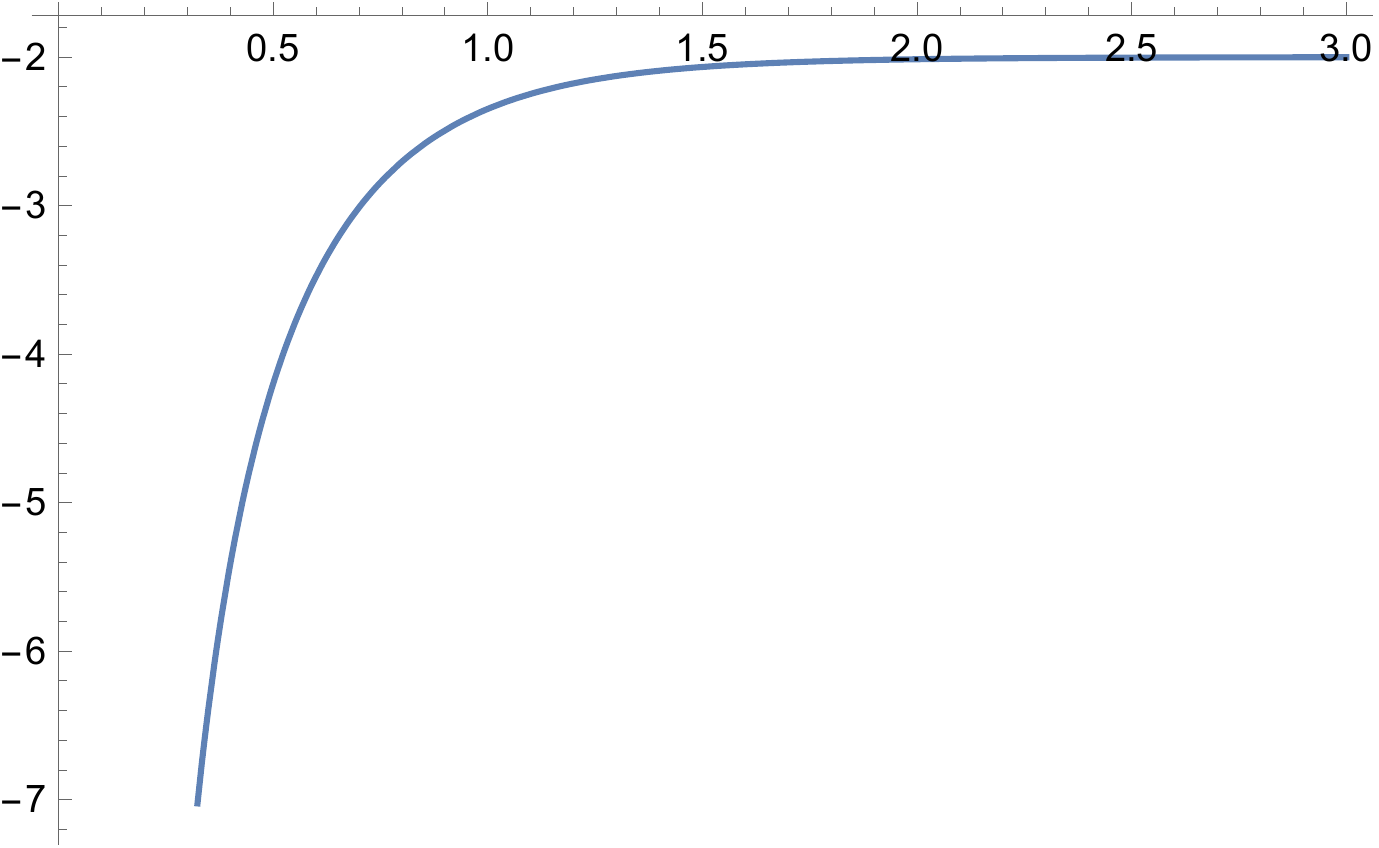} 
\end{tabular}
\caption{Left figure $H'/\lambda E-1$  for $\lambda=1$.}\label{figdiffonegzero}
\end{figure}

Not surprisingly this is the same and solution for $E$ and $B$ in the Maxwell +CS system and 
hence it is not a solution of the BI case. Indeed 
\be
H'=\lambda E
\ee
is not fulfilled. This can be seen in figure \ref{figdiffonegzero}.

If indeed these configurations of $\tilde D ,B$ are solutions of the full system of equations, then it is easy to check, 
using (\ref{HDB}), that the corresponding energy when we take $d=0$, unlike the Maxwell case, is finite.

\subsection{Attempt 2}

Another attempt which is inspired by the passage of the electric field from Maxwell theory to the BI one, namely
\be
E\sim \frac{1}{r}\rightarrow  \frac{1}{\sqrt{1+ r^2}}\;,
\ee
takes the form
\be\label{Datt}
\tilde D \sim K_1[ \lambda\rho ] \rightarrow K_1[ \lambda\rho ]/\sqrt{1+ K_1[ \lambda\rho ]^2]}\;,
\ee
and similarly 
\be
B \sim K_1[ \lambda\rho ] \rightarrow K_0[ \lambda\rho ]/\sqrt{1+ K_0[ \lambda\rho ]^2}.
\ee

This means that we take 
\be
f(\rho\lambda)=\frac{}{\sqrt{1+ K_1(\rho\lambda)^2}} \qquad f(\rho\lambda)+g(\rho\lambda)\frac{}{\sqrt{1+ K_0(\rho\lambda)^2}}.
\ee

\begin{figure}[t!]
\begin{tabular} {c}
\includegraphics[width=6cm]{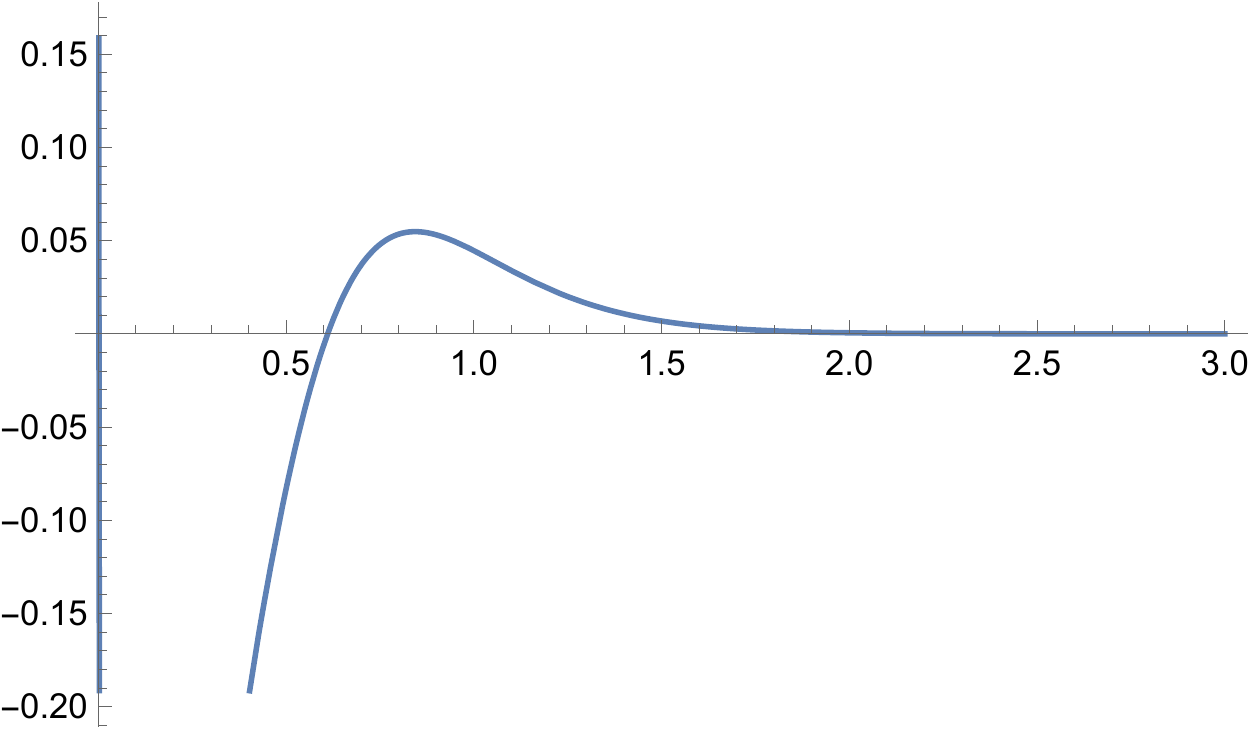}\includegraphics[width=6cm]{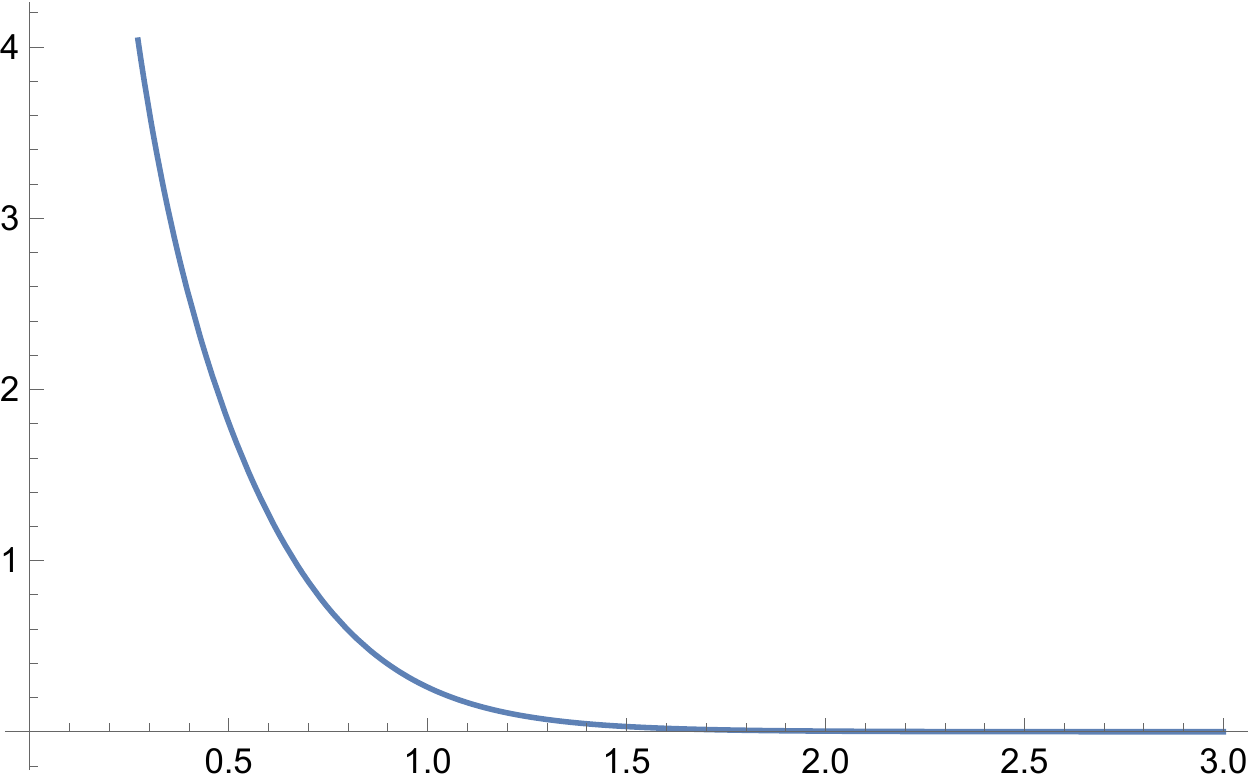}
\end{tabular}
\caption{Left figure $H'-\lambda E$. Right figure $D'+D\rho-\lambda B$ for $\lambda=1$.}\label{figdif}
\end{figure}

In this case, using the constitutive relations, we find that the difference between 
$H'[\lambda\rho]-\lambda E[\lambda\rho]$ and  $D[\lambda\rho]' +D[\lambda\rho]-\lambda B[\lambda\rho]$
is very small, apart from the region around $\lambda\rho\sim 0$, as can be  seen in the figure \ref{figdif}.

This is not surprising, since the BI starts to deviate from Maxwell when $\lambda\rho\sim 1$.

\begin{figure}[t!]
\begin{tabular} {c}
\includegraphics[width=10cm]{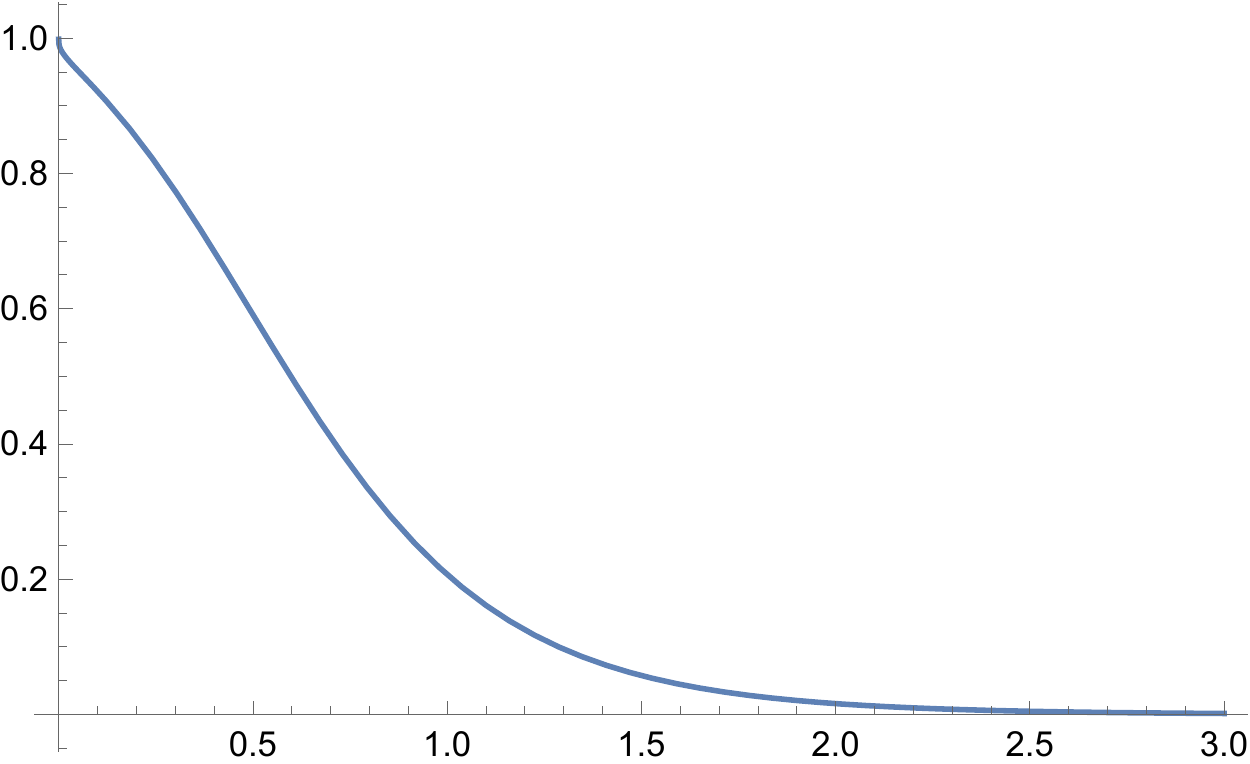} 
\end{tabular}
\caption{ The energy as a function of $\lambda\rho$ for $g=b=\lambda=1$.}\label{figenr}
\end{figure}

The energy density associated with this configuration, following (\ref{HDB}), is drawn in figure \ref{figenr}.

It is obvious from this figure that the total energy is indeed finite.
The question is whether the correction to this configuration that yields a solution of the system will preserve this property.

\subsection{Attempt 3}

A third attempt of finding an analytic solution is as follows. We start with the ansatz  for $D$ of above (\ref{Datt}).
We then deterimine $B $ using (\ref{eom}) and get
\be
B=\frac{K_1(r){}^3-r K_0(r)}{r \left(K_1(r){}^2+1\right){}^{3/2}}.
\ee

\begin{figure}[t!]
\begin{tabular} {c}
\includegraphics[width=10cm]{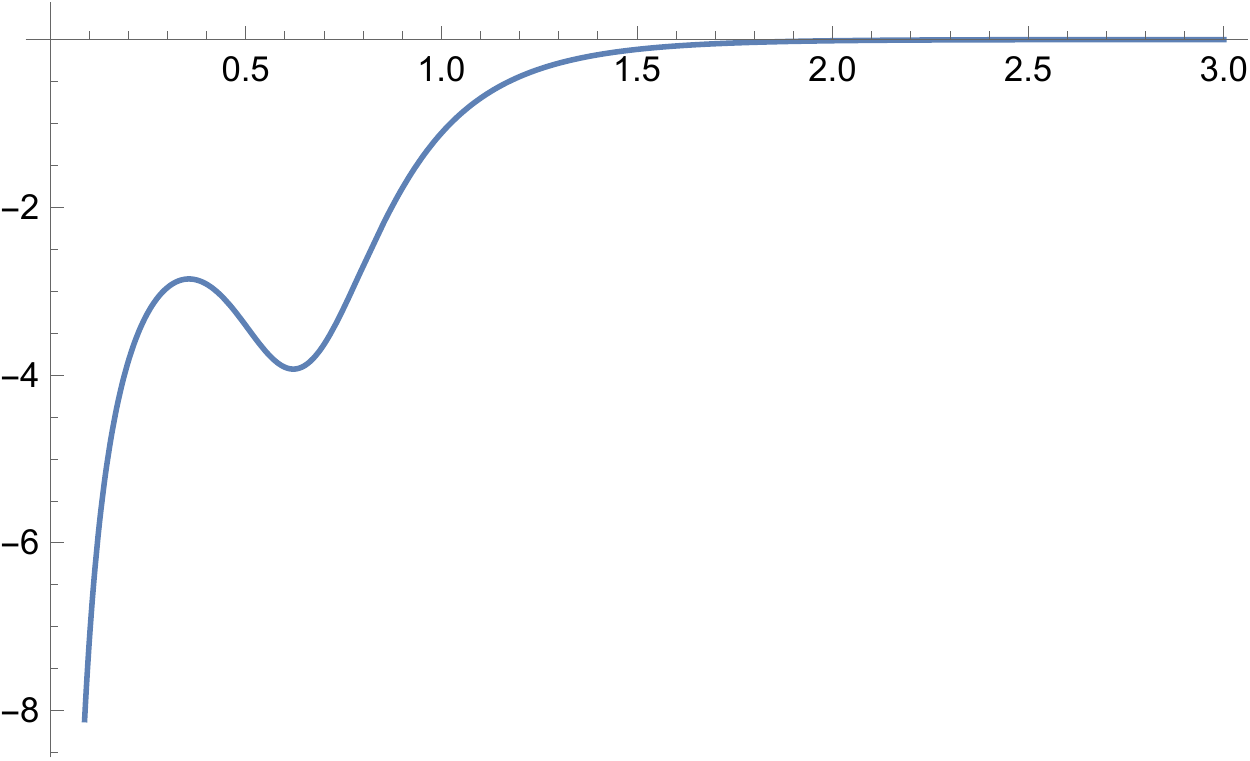} 
\end{tabular}
\caption{$H'-\lambda E$ for attempt 3 with  $\lambda=1$.}\label{figdif3}
\end{figure}

We then determine $E$ and $H$ using the constitutive relations and check again whether the other eom 
$H'-\lambda E$ is obeyed. Again it is obeyed apart from the region of $\rho\sim 0$ as can be seen in figure \ref{figdif3}.
 
 \subsection{Attempt 4}

Another attempt is to use  (\ref{Datt}) for $B[ \lambda\rho ]$, but for $B[ \lambda\rho ]$ we take
\be
B \sim K_1[ \lambda\rho ] \rightarrow K_0[ \lambda\rho ]/\sqrt{1+ K_0[ \lambda\rho ]^2}.
\ee

\begin{figure}[t!]
\begin{tabular} {c}
\includegraphics[width=6cm]{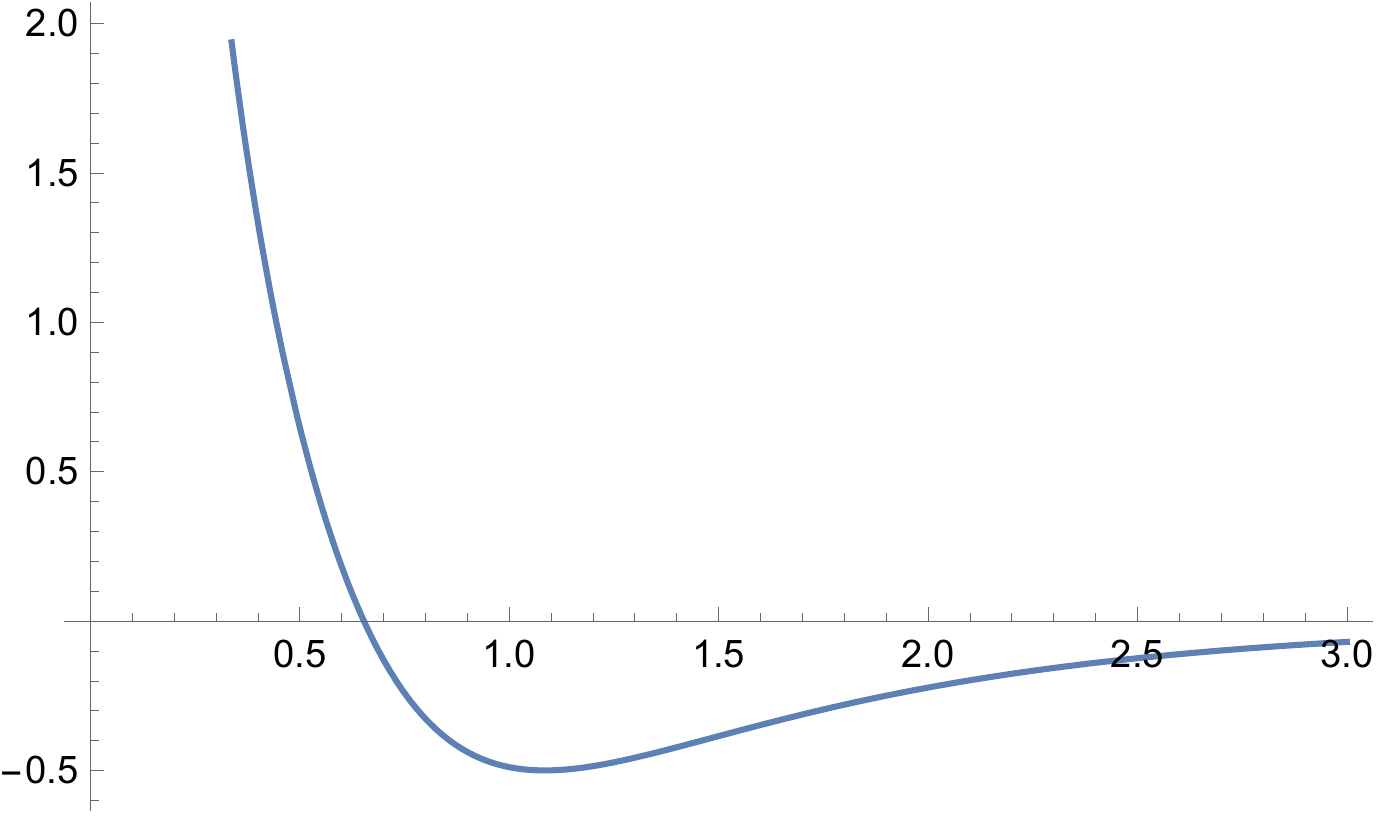}\includegraphics[width=6cm]{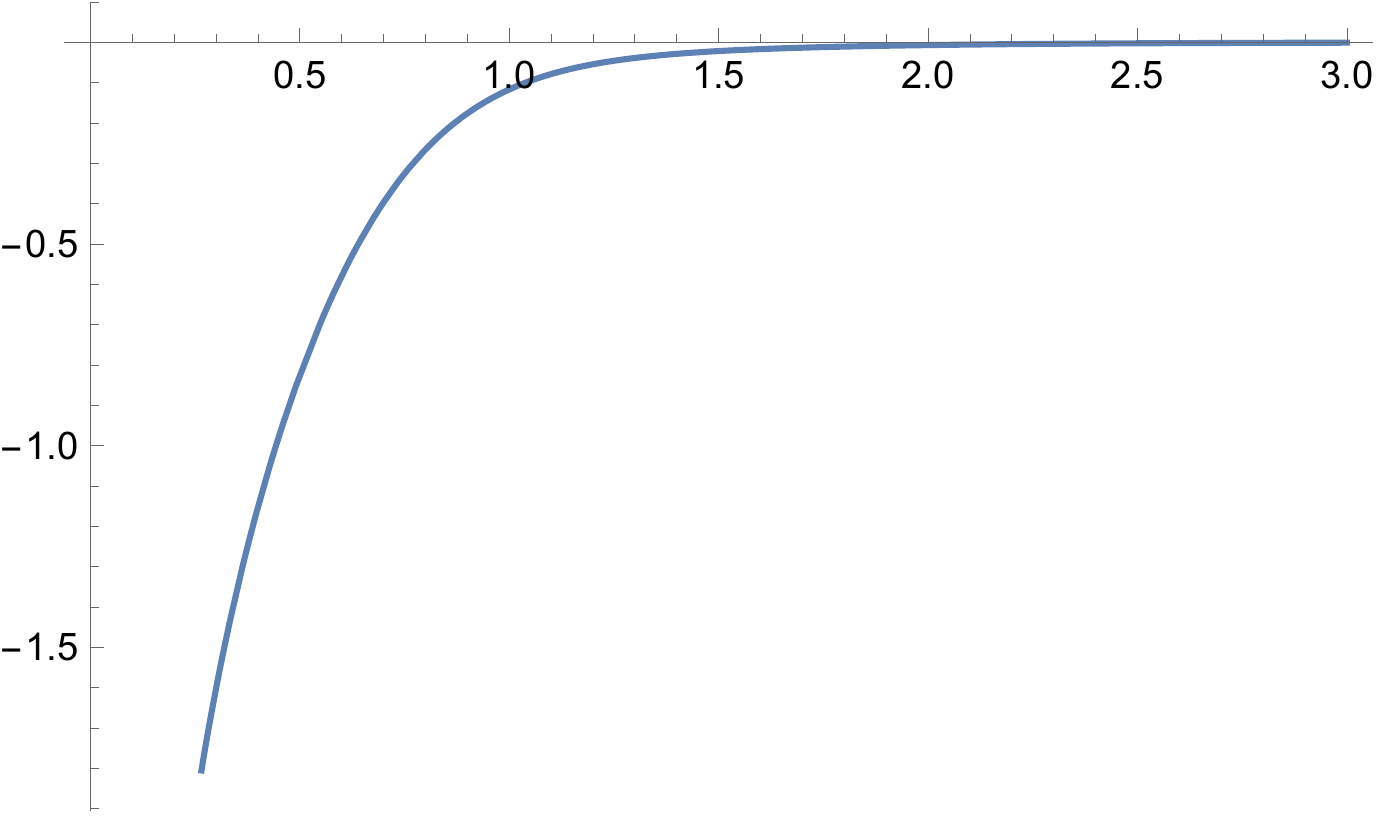}
\end{tabular}
\caption{Right figure $H'-\lambda E$. Left figure $D'+D\rho-\lambda B$ for $\lambda=1$.}\label{figdifc}
\end{figure}

In this case  the configurations are a reasonable approximation for the solutions of the eom for large $\rho$ but deviate 
in the region of small $\rho$, as can be seen in figure \ref{figdifc}, do no solve exactly the equations of motion 
in the region of small $\rho$.

To conclude, we see that we could not find an analytic solution.

\bibliography{CSBI}
\bibliographystyle{utphys}

\end{document}